\documentclass[prx,aps,amssymb,twocolumn,showpacs,superscriptaddress,footinbib]{revtex4-1}
\usepackage{graphicx}
\usepackage{amsmath,amsfonts,amssymb}

\begin{document}

\title{Quantum fluctuations of the current in a tunnel junction at optical frequencies}

\author{P. F\'{e}vrier}
\affiliation{Laboratoire de Physique des Solides, CNRS, Univ. Paris-Sud, Universit\'{e} Paris-Saclay, 91405 Orsay Cedex, France}
\author{J. Gabelli}
\affiliation{Laboratoire de Physique des Solides, CNRS, Univ. Paris-Sud, Universit\'{e} Paris-Saclay, 91405 Orsay Cedex, France}

\date{\today}

\begin{abstract}
We have investigated the mechanism at the origin of the infra-red radiation emitted by a biased tunnel junction by detecting photons at frequencies $\nu<eV/h$. To address this regime, the bias voltage $V$ exceeds one volt and the potential profile of the tunnel barrier is driven far from its equilibrium state. As a consequence, the $I(V)$ characteristic of the junction is strongly nonlinear. At optical frequencies, the transport through the junction cannot be simply expressed in term of the dc current and the current fluctuations are no longer described by the fluctuation-dissipation relation. Taking into account the energy and voltage dependence of the transmission of the tunnel junction in a Landauer-B\"{u}ttiker scattering approach, we experimentally demonstrate that the photon emission results from the fluctuations of the current inside the tunneling barrier.
\end{abstract}

\maketitle



\begin{figure}
\begin{center}
\includegraphics[width=0.9\linewidth]{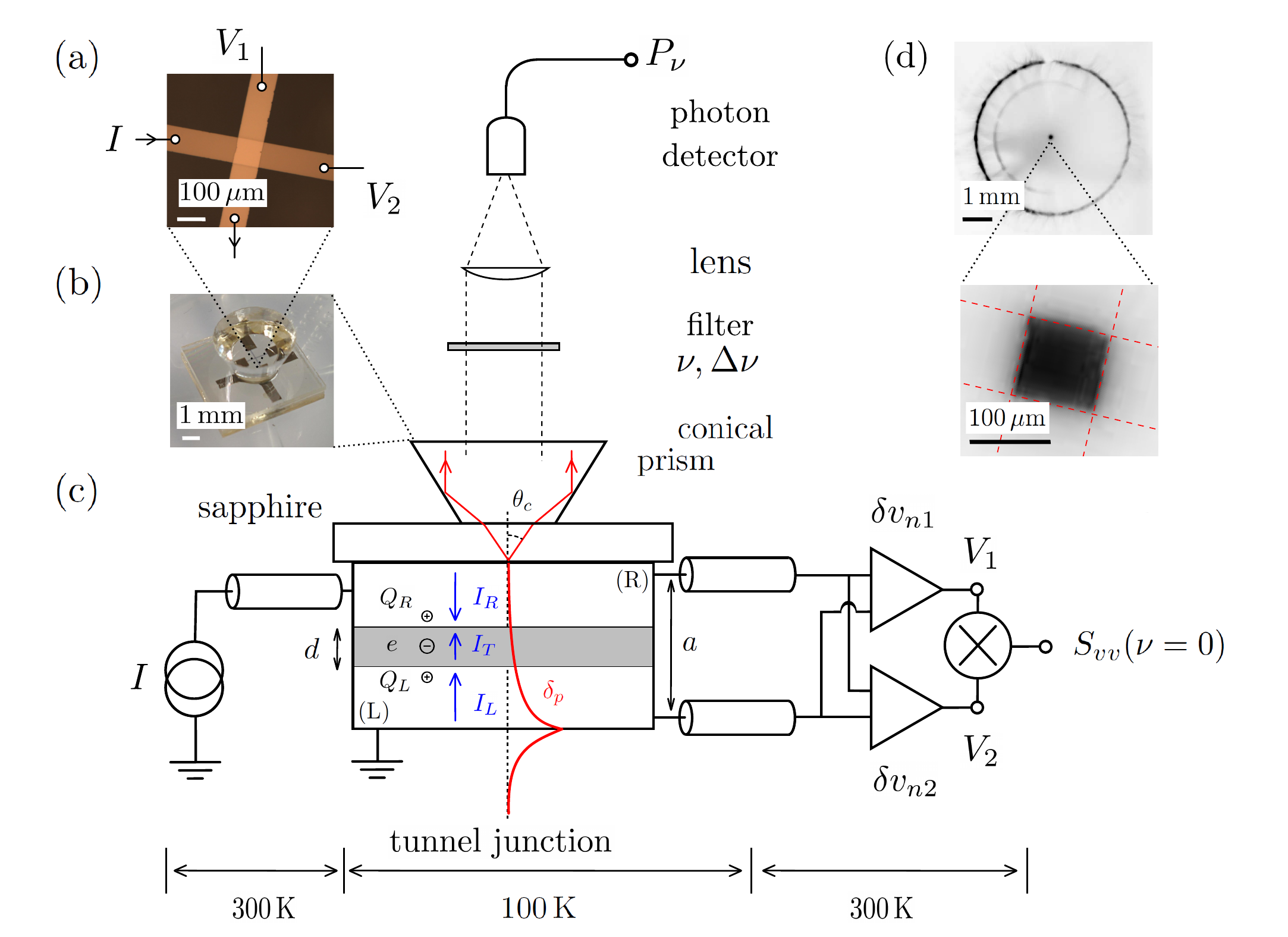}
\end{center}
\vspace{-0.5cm}
\caption{(a) Optical micrograph of the metallic cross-junction. (b) Optical picture of the sample. The conical prism allows to collect photons thanks to a total internal reflection.  (c) Schematic of the experimental setup. (d) Emitted light from the tunnel junction ($I=1.7 \, \mathrm{mA}$) directly observed with a sensitive camera in the spectral range $0.4 - 1 \, \mu \mathrm{m}$.
\label{fig1}}
\end{figure}

Fluctuations of the current in a conductor give rise to electromagnetic radiation. In the free space at thermal equilibrium, the radiated spectral power is described by Planck's law and is a direct consequence of the \emph{fluctuation-dissipation theorem} (FDT): the thermal fluctuating currents in the conductor generate an electromagnetic field related to the dissipation in the conductor through its resistivity \cite{Nyquist_PhysRev.32.110,Callen_PhysRev.83.34}. Besides thermal fluctuations, conductors can experience another fundamental source of current fluctuations, the so-called \emph{shot noise}. A natural question arise : can the black-body law be generalized to current-biased conductors? If such a generalization exists, it should  particularly be observed in conductors exhibiting  Poissonian shot noise like tunnel junctions. Even though broadband light emitted from metallic tunnel junctions was first observed in the late 70's by Lamb and McCarthy \cite{Lamb_PhysRevLett.37.923}, no general relation has been established so far between the emission spectrum at optical frequencies and the electronic transport through the junction \cite{Kirtley81,Hanisch94}. Following the Nyquist argument \cite{Nyquist_PhysRev.32.110}, the radiated spectral power $P_{\nu}$ emitted by a planar tunnel junction can be expressed in terms of the current noise spectral density $S_{ii}$ and a radiation impedance $\mathcal{R}(\nu)$ standing for the coupling between the tunneling currents in the conductor and the far field radiating electromagnetic modes:
\small
\begin{equation}\label{eq.R}
P_{\nu}=\mathcal{R}(\nu) S_{ii}
\end{equation}
\normalsize
\noindent In the case of a tunnel junction at thermal equilibrium, the FDT gives $S_{ii}(h\nu)=2G \, h\nu \, N(h\nu)$ where $N(\epsilon)=1/((\mathrm{exp}(\epsilon/k_BT)-1)$ denotes the Bose-Einstein distribution and $G$ the dc conductance  of the junction. For a dc-polarized tunnel junction, the FDT has been generalized to an expression which is usually referred to as a fluctuation dissipation relation (FDR) \cite{Rogovin74,Lee_PhysRevB.53.7383,Sukhorukov_PhysRevB.63.125315,Roussel_PhysRevB.93.045102}:
\small
\begin{equation}\label{eq.FDR}
\begin{split}
S_{ii}^{(FDR)}(eV,h\nu)= & e \left\{(N(eV-h\nu)+1) \, I(V-h\nu/e) \right.\\
 & \hspace{0.7cm} \left.+N(eV+h\nu) \, I(V+h\nu/e) \right\}
\end{split}
\end{equation}
\normalsize
\noindent where $I(V)$ is the dc characteristic of the voltage-biased tunnel junction. This prediction is in quantitative agreement with experiments in the microwave regime in a linear tunnel junction \cite{Basset_PhysRevLett.105.166801,Gabelli_PhysRevLett.100.026601} or in a tunnel junction showing non-linear features of dynamical Coulomb blockade \cite{Parlavecchio_PhysRevLett.114.126801}. Although this FDR is universal at zero frequency and can be deduced from a general fluctuation theorem \cite{SM,PhysRevLett.71.2401,PhysRevB.75.155316}, we show in this letter that it breaks down at optical frequencies ($\lambda \sim 1 \, \mu\mathrm{m}$).  Eq.~(\ref{eq.FDR}) is indeed based on a perturbation theory applied to a model transfer Hamiltonian \cite{Bardeen61,Roussel_PhysRevB.93.045102} and cannot stand when the bias voltage is comparable with the tunneling barrier height. First, the tunneling barrier is modified by the bias voltage leading to an intrinsic non-linearity  conductance. Second, the spectral noise density measured at optical frequencies probes current correlations on a time scale $\tau \sim 3 \times 10^{-15} \, \mathrm{s}$ on the order of the time for an electron to cross the barrier \cite{SM,Buttiker_PhysRevLett.49.1739}. The photon emission is then a ``snapshot'' of the tunneling event and requires a microscopic description of the charge transfer inside the tunneling barrier. Our experiments not only shed light on the origin of light emission by tunnel junctions, but also extend the concepts of low-energy electronic transport to a few $eV$ and explore the new regime of finite frequency quantum noise in nonlinear transport. The letter is organized as follows: (i) we define the transport in a tunnel junction in the far-from-equilibrium regime (FFER). (ii) We describe the experimental setup (FIG.~\ref{fig1}). (iii) We experimentally show that the FDR holds on in the FFER at zero frequency proving the validity of the tunneling limit. (iv) We use a Landauer-B\"{u}ttiker (LB) approach based on elastic tunneling processes to quantitatively describe the noise spectral density in the optical spectral range.

\begin{figure}
\begin{center}
\includegraphics[width=0.9\linewidth]{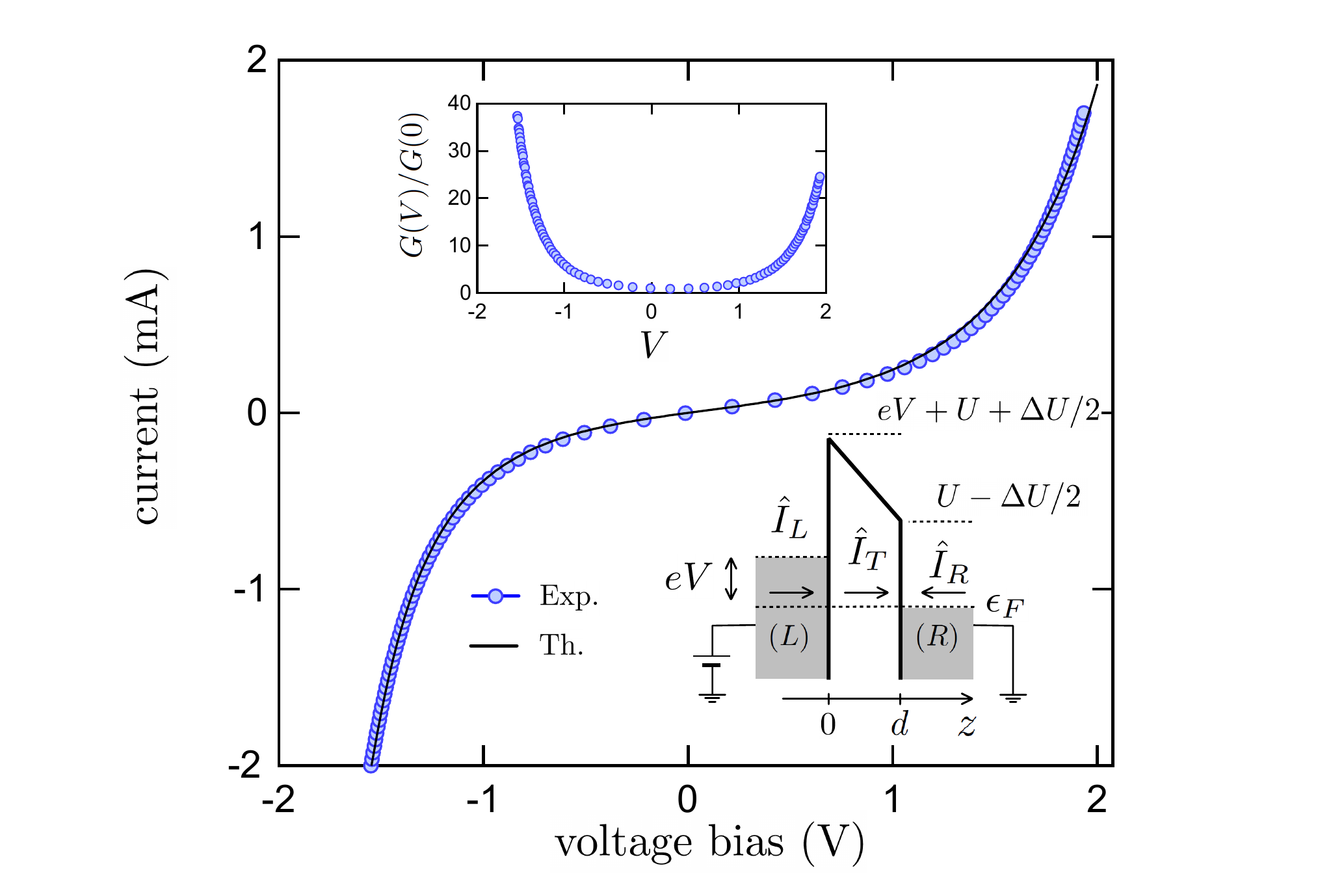}
\end{center}
\vspace{-0.5cm}
\caption{$I(V)$ characteristic of the tunnel junction. Symbols are experimental data and solid lines are theoretical expectations of Eq.~(\ref{eq.IV}) using the WKB approximation on a trapezoidal barrier characterized by a mean height $U$ and an asymmetry $\Delta U$. \textit{Upper inset}: differential conductance \textit{vs.} voltage at low bias. \textit{Lower inset}: schematic of the trapezoidal barrier modified by a bias voltage $V$. \label{fig2}}
\end{figure}

\vspace{0.2cm}\noindent\emph{Nonlinear tunneling transport.} The FFER is achieved when the applied bias voltage is of the order of the tunnel barrier height $U$. In this regime, without a careful study of the Coulomb interactions in the tunnel barrier, gauge invariance (invariance of the current under a global voltage shift applied on both electrodes) is not systematically satisfied  \cite{Christen_EuroPhys96,Blanter2001}. It is indeed necessary to determine the electrical potential which depends on the applied bias voltage and the possible charge accumulation in the conductor. The transmission $\mathcal{T}$ of the barrier is thus necessarily \emph{energy} and \emph{voltage} dependent and the $I(V)$ characteristic is expressed according to the Landauer-B\"{u}ttiker formula as:
\small
\begin{equation}\label{eq.IV}
I(V)=\frac{2e}{h}\int d\epsilon\,\mathcal{T}(\epsilon,eV) \left\{f(\epsilon-eV)-f(\epsilon)\right\}
\end{equation}
\normalsize \noindent where $f(\epsilon)=1/\left(1+\mathrm{exp}((\epsilon-\epsilon_F)/k_BT)\right)$ is the Fermi-Dirac distribution with $\epsilon_F$ the Fermi energy. In the tunneling limit, the voltage dependence of  $\mathcal{T}$ can be deduced from the Wentzel-Kramers-Brillouin (WKB) approximation by considering a total potential including the potential barrier $U(z)$ and the biasing energy $U_{bias}(z,V)=eV(1-z/d)$ as depicted in FIG.~\ref{fig2}\cite{Simmons64}. It is worth  emphasizing  that the biasing energy is essential to explain non-symmetric $I(V)$ characteristics as shown in FIG.~\ref{fig2}. We now consider the current fluctuations characterized by the non-symmetrized spectral noise density  $S_{\alpha,\beta}=\langle \hat{I}_{\alpha}(-\nu)\hat{I}_{\beta}(\nu)\rangle \Delta f$  where $\hat{I}_{\alpha}(\nu)$ is the Fourier component of the current operator measured in the electrode $\alpha=L,R$  and $\Delta f$ the measurement bandwidth. For $\nu >0$, this quantity refers to the emission quantum noise which is measured in a passive detection scheme such as the photon detector used here \cite{Lesovik97,Blanter2001}. Using the scattering LB approach for a single quantum channel of conduction in the tunneling limit ($\mathcal{T} \ll 1$), we get for $\alpha \neq \beta$ \cite{Blanter2001}:
\small
\begin{subequations}
\label{allequations}
\begin{align}
S_{\alpha \alpha}(eV,h\nu)=&\frac{e^2}{h}\int d\epsilon \left\{\mathcal{T}(\epsilon{-}h\nu,eV) f_{\alpha}(\epsilon)\right. \nonumber\\
& \hspace{-2cm} \left. \times (1{-}f_{\beta}(\epsilon{-}h\nu)){+}\mathcal{T}(\epsilon,eV) f_{\beta}(\epsilon)(1{-}f_{\alpha}(\epsilon{-}h\nu)) \right\}\label{eq.SLL}\\
S_{\alpha \beta}(eV,h\nu)=&-\frac{e^2}{h}\int d\epsilon \sqrt{\mathcal{T}(\epsilon,eV)\mathcal{T}(\epsilon{-}h\nu,eV)}\nonumber \\
& \hspace{-2cm} \times \left\{ f_{\alpha}(\epsilon)(1{-}f_{\beta}(\epsilon{-}h\nu)){+}f_{\beta}(\epsilon)(1{-}f_{\alpha}(\epsilon{-}h\nu))\right\}\label{eq.SLR}
\end{align}
\end{subequations}
\normalsize
\noindent where $f_L(\epsilon)=f(\epsilon-eV)$ and $f_R(\epsilon)=f(\epsilon)$. In the zero-frequency limit, a straightforward calculation leads to $S_{LL}=S_{RR}=-S_{LR}=S_{ii}^{(FDR)}$ and the FDR holds even in the nonlinear regime. However, at finite frequency, the energy dependence of the transmission  $\mathcal{T}$ leads to a \emph{charge accumulation in the barrier} and the noise spectral density depends on the electrode where it is evaluated ($S_{LL} \neq S_{RR} \neq -S_{LR}$) \cite{Blanter2001,Zamoum_PhysRevB.93.235449}. Because of the screening of the electromagnetic field in the metallic electrodes, the coupling is expected to be dominant in the insulating barrier and requires the determination of the tunneling current to evaluate the  radiation impedance. Although it should be necessary to solve the coupled system of Schr\"{o}dinger and Poisson equations to calculate the tunneling current $\hat{I}_T$, the screening in metallic electrodes enables a simple description of $\hat{I}_T$. The bare electron inside the tunneling barrier induces a polarization charge $-e(1-z/d)$ and $-ez/d$ in the left and right electrodes respectively (see FIG.~\ref{fig2}). We can thus assume that the charge accumulation on the surface of the electrodes is equal in average during the tunneling event: $\hat{Q}_L=\hat{Q}_R=\hat{Q}$ with $\langle \hat{Q} \rangle=-e/2$. The continuity equation $d\hat{Q}/dt=\hat{I}_L-\hat{I}_T=\hat{I}_T+\hat{I}_R$ then implies $\hat{I}_T=(\hat{I}_L-\hat{I}_R)/2$ with the conventional direction of the current (FIG.~\ref{fig1}(c)) \cite{PhysRevB.88.045413}. The current noise spectral density $S_{ii}=\langle \hat{I}_{T}(-\nu)\hat{I}_{T}(\nu)\rangle \Delta f$ in  Eq.~(\ref{eq.R}) is then given by:
\small
\begin{equation}\label{eq.Sii}
S_{ii}(eV,h\nu) = \frac{1}{4} \left(S_{LL}+S_{RR}-2S_{LR} \right)
\end{equation}
\normalsize
\noindent while the radiation impedance $\mathcal{R}(\nu)$ is associated with the leakage of the surface plasmon polariton (SPP) mode in the substrate (FIG~\ref{fig1}(c)). Under these conditions, the FDR cannot be satisfied anymore and the radiated spectral power $P_{\nu}$ measured by the photon detector is a linear combination of $S_{LL}$, $S_{RR}$ and $S_{LR}$ given by the coupling between the current fluctuations and the electric field in the junction. It should be stressed that the expression of $\hat{I}_T$ account for all the effects of Coulomb interactions.

\vspace{0.2cm}\noindent\emph{Experimental setup.} Our experimental setup is shown in FIG.~\ref{fig1}. Electronic and optical measurements are performed in a cryogenic environment at $T \sim 100 \,\mathrm{K}$ to prevent junction breakdown and to reduce the thermal noise on the infrared photon-detector. The sample is a $100 \times 100 \, \mu \mathrm{m}^2$ planar $Al/AlO_x/Al$ tunnel junction deposited on a sapphire substrate (FIG.~\ref{fig1}(a)). Because of the layered structure of the junction, the electromagnetic modes are localized in the junction and consequently should not radiate in the free space. However, the total thickness of the junction $a \sim 10 \, \mathrm{nm}$  is smaller than the penetration depth of the SPP in the metal: $\delta_p=c/\omega_p \simeq 13 \, \mathrm{nm}$ with $\omega_p=14.7 \, \mathrm{eV}$ the plasma frequency of aluminum. It then allows the coupling between the SPP mode localized at the interface electrode/vacuum (FIG~\ref{fig1}(c)) and the propagating mode in the substrate \cite{SM}. This corresponds to the Kretschmann configuration where the coupling appears at a specific angle $\theta_p \simeq \arcsin (1/n) \sim 35^{\circ}$ where $n$ stands for the refractive index of sapphire \cite{Kretschmann72}. We use total internal reflection in a conical prism to collect the emitted photons (see FIG.~\ref{fig1}(b,c)). The current noise $S_{ii}(eV,h\nu)$ at optical frequency $\nu$ is measured at two different frequencies corresponding to the wavelengths $\lambda=c/\nu=0.9 \pm 0.02 \, \mu \mathrm{m}$ and $1.3\pm 0.015 \, \mu \mathrm{m}$. The current noise at zero frequency $S_{ii}(eV,h\nu=0)$  is measured with a standard cross-correlation technique \cite{SM}.

\vspace{0.2cm}\noindent\emph{Electrical properties - Noise measurement at zero frequency.} At high voltage the $I(V)$ characteristic shown in FIG.~\ref{fig2} exhibits a strong nonlinearity: the differential resistance varies by more than one order of magnitude going from $6 \,\mathrm{k}\Omega$ at low bias to $150 \,\Omega$ at high bias. From the theoretical expectation of Eq.~(\ref{eq.IV}) using the WKB approximation to evaluate the transmission of the tunnel junction, we estimate the mean barrier height $U \sim 2.7 \, eV $, its asymmetry $\Delta U \sim 2.9 \, eV$ and its thickness $d \sim 2 \, \mathrm{nm}$ (lower inset of FIG.~\ref{fig2}). The thickness is in agreement with the capacitance of the junction $\sim 0.5\, \mathrm{nF}$.
\begin{figure}
\begin{center}
\includegraphics[width=0.9\linewidth]{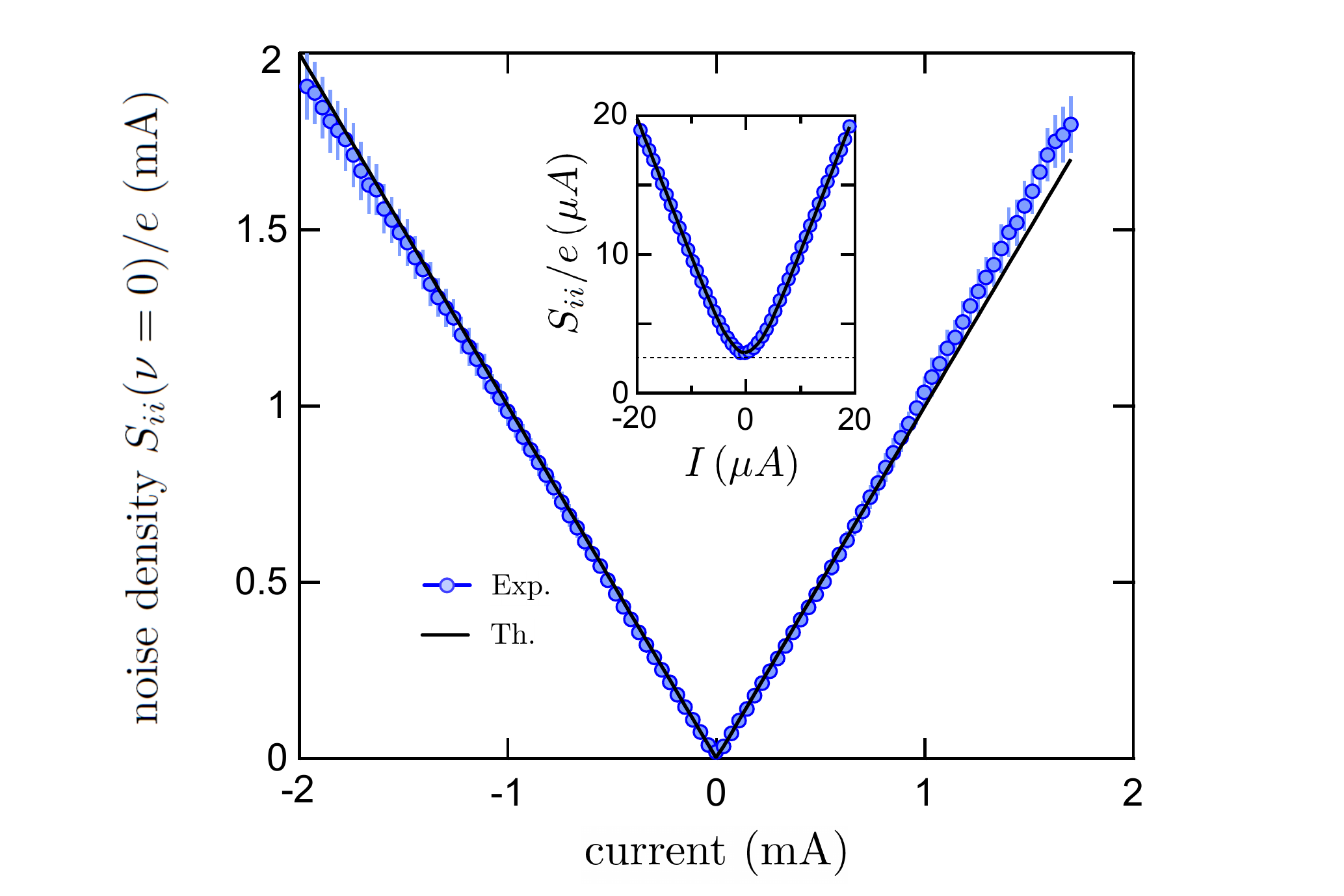}
\end{center}
\vspace{-0.5cm}
\caption{(a) Electronic  shot noise of the tunnel junction measured in the bandwidth $20-100 \, \mathrm{kHz}$. \textit{Inset}: Zoom at low voltage bias. Solid lines correspond to theoretical expectation of Eq.~(\ref{eq.Sii0}) with $T=100 \, \mathrm{K}$.\label{fig3}}
\end{figure}
\noindent  Current fluctuations at zero frequency $S_{ii}(eV,h\nu=0)$ are measured with low noise voltage amplifiers giving access to voltage fluctuation $S_{vv}=g\left(|Z_{setup}(eV)|^2 S_{ii}+S_{vv,setup}(eV)\right)$ where $g$ is the global gain of the amplifier chain, $Z_{setup}$ is the transimpedance of the measurement setup and $S_{vv,setup}$ its excess noise. Because of the large variation of the tunneling resistance, a careful calibration is required to extract the current noise $S_{ii}$. The voltage-dependent transimpedance $Z_{setup}$ and the excess noise $S_{vv,setup}$ are determined by using an external noise source while $g$ is deduced from the measurement of the shot noise in the linear regime \cite{SM}. FIG.~\ref{fig3} shows the current noise $S_{ii}$ in the FFER. Although the tunnel resistance is strongly nonlinear, $S_{ii}$ clearly satisfies the FDR at zero frequency:
\small
\begin{equation}\label{eq.Sii0}
S_{ii}(eV,h\nu=0)= \frac{eI(V)}{\tanh \left( eV/2k_BT \right)}
\end{equation}
\normalsize
\noindent  In the high bias limit $eV \gg k_BT$, the current noise is then linearly proportional to the dc current which is a signature of shot noise. It confirms that electronic transport through the junction operates in the \emph{tunneling  limit} at high voltage bias ruling out the presence of pinholes in the barrier. We notice systematic errors at high positive bias. They cannot be attributed to Joule heating since they should also be observed for negative bias. The fact that calibration is off by $\sim 10 \%$ is attributed to parasitic capacitances of the measurement setup which are not included in $Z_{setup}$.

\begin{figure}
\begin{center}
\includegraphics[width=0.9\linewidth]{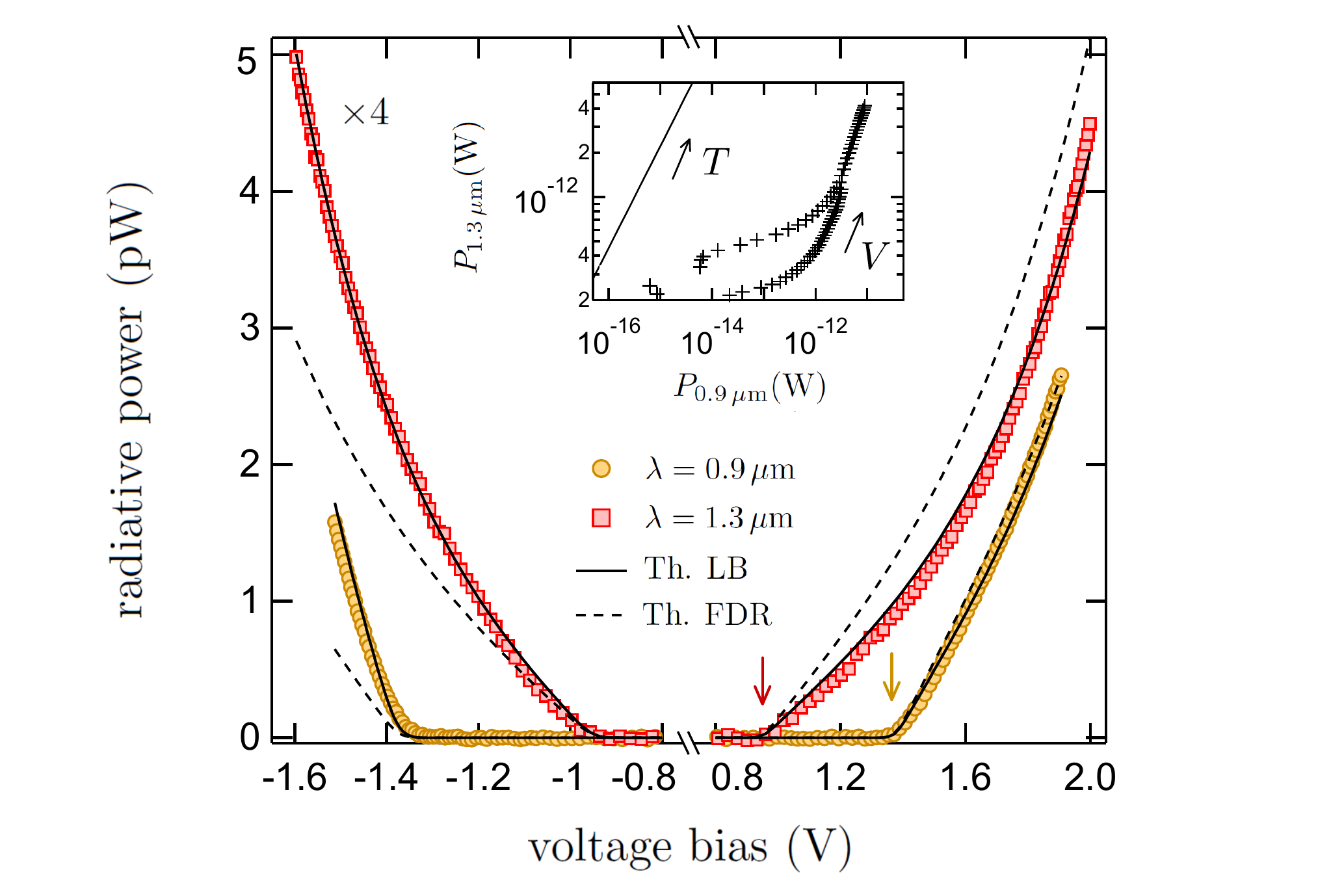}
\end{center}
\vspace{-0.5cm}
\caption{(a) Radiated power as a function of voltage bias $V$ in the configuration depicted in FIG.~\ref{fig1}. Markers correspond to data recorded at different wavelengths. Data at $\lambda=1.3 \,\mu \mathrm{m}$ (red squares) have been multiplied by a factor 4 for clarity. Vertical arrows define the voltage cut-off $eV=hc/\lambda$. Solid lines and dashed lines are  theoretical  expectations from  Eq.~(\ref{eq.Sii}) and Eq.~(\ref{eq.FDR}) respectively. Horizontal axis has been split for clarity. \textit{Inset}: relationship between the light power at two different wavelengths with increasing bias voltage. Solid line corresponds to the black-body law with increasing temperature.\label{fig4}}
\end{figure}
\noindent

\vspace{0.2cm}\noindent\emph{Light emission - Noise measurement at optical frequency.} FIG.~\ref{fig1}(d) shows an image of the light emission pattern from the tunnel junction when the camera is focused on the conical prism. In the center, a small amount of light comes directly from the tunnel junction (zoom in FIG.~\ref{fig1}(d)). This is due to surface roughness of electrodes allowing SPP scattering at the surface of the upper electrode \cite{Hanisch94}. The homogenous light intensity indicates that electron to photon conversion in the tunnel junction is also homogenous over the surface of the junction. However, the bright ring in FIG.~\ref{fig1}(d) reveals that more than $98 \%$ of the light is emitted at the specific angle $\theta_{p}$ as expected in the Kretschmann configuration. The light power $P_{\nu} \Delta \nu$ is plotted as a function of the voltage bias for two different wavelengths $\lambda$ in FIG.~\ref{fig4}. Inset of FIG.~\ref{fig4} displays the relationship between the light power at the two wavelengths on a log-log plot \cite{gabelli10}. Data points do not fit the black-body radiation law (solid line in inset) and, as previously mentioned, the Joule heating cannot be responsible for the observed photon emission. The light power exhibits a voltage cross-over at $eV=hc/\lambda$: electrons crossing the tunnel junction relax their energy by emitting photons at frequency $\nu \leq eV/h$. This cross-over is predicted both by the FDR and the LB theories. However, our data clearly disagree with the FDR (dashed line in FIG.~\ref{fig4}) and are in very good agreement with the LB relation of Eq.~(\ref{eq.Sii}) (solid line in FIG.~\ref{fig4}). The LB approach enables us to understand the dependence on the bias polarity of the light emission which has already been observed but not explained \cite{Kirtley81,Kirtley83,Hanisch94}. It also allows one to extract the radiation impedance according to Eq.~(\ref{eq.R}). This gives $\mathcal{R}(\lambda=0.9 \, \mu \mathrm{m})=2.5 \, \mathrm{m} \Omega$ and $\mathcal{R}(\lambda=1.3 \, \mu \mathrm{m})=2.3 \, \mathrm{m} \Omega$ about a factor four higher than our rough estimation in the limit $\nu/\nu_p\ll 1$ \cite{Laks79,Ushioda86}:
\small
\begin{equation}\label{eq.Restimat}
\mathcal{R}(\nu)=\frac{1}{\beta n^5} \left(\frac{d}{\delta_p} \right)^2 \left(\frac{\nu}{\nu_p} \right)^3  Z_{vac}  \underset{\lambda=1 \mu \mathrm{m}}{\sim}   0.5 \, \mathrm{m} \Omega
\end{equation}
\normalsize
\noindent where $\beta=\tanh(a/\delta_p)\simeq 0.69$, $\nu_p=\omega_p/2\pi$, $n=1.75$ is the refractive index of sapphire and the alumina dielectric barrier, $d \simeq 2 \, \mathrm{nm}$ is the thickness of the barrier and $Z_{vac} \simeq 376 \, \Omega$ is the vacuum impedance. This under-estimation can be attributed to the approximative values of the thickness and the refractive index of the dielectric barrier but also to the interband transition at $\lambda_{inter}=0.825 \, \mu \mathrm{m}$ in aluminum. We assume here that the coupling between the current fluctuations and the electric field takes place in the insulating barrier. This is justified by the screening of the electric field in the metal. If we only consider the coupling in the electrodes, we indeed expect a radiation impedance in the $\mu \Omega$ range, three orders of magnitude smaller than the observed one \cite{SM}. However, the radiation impedance in the $\mathrm{m}\Omega$ range is rather small and appears as a central quantity in the understanding of the small emission light efficiency of tunnel junctions. This lead us to redefine the efficiency with respect to the dissipated Joule power: $\eta=\int_0^{+\infty} P_{\nu}d\nu/(V\times I) \sim 4 \times 10^{-8}$. According to this definition, we can show that the efficiency is now directly related to the radiation impedance: $\eta \sim \eta_0 \mathcal{R}(eV/h)/R_K$ where $R_K=h/e^2 \simeq 25.8 \, \mathrm{k}\Omega$ is the quantum of resistance and $\eta_0 \simeq 0.047$ is a constant slightly dependent on the details of the barrier \cite{SM}. We emphasize that this definition contrasts with the usual one which is given by the electron-to-photon conversion rate. We find the former more appropriate since it reflects the fact that, in metallic tunnel junctions, electrons with energy smaller than bias voltage can contribute to the current. In fact, unlike semiconductors, the lack of a band gap in metals indeed implies that each electron crossing the barrier emits a bunch of photons in a spectral range $0<\nu < eV/h$ with a radiated spectral power proportional to the current. The emitted light power is then proportional to the Joule power $V\times I$.

\vspace{0.2cm}\noindent\emph{Discission.} The photon emission in a tunnel junction is usually attributed to the spontaneous emission in the barrier by inelastic electron tunneling \cite{Persson92,Berndt91,Parzefall15}. However, it is worth noting that the LB approach which is used here only describes \emph{elastic tunneling processes}. In this description, the energy relaxation formally takes place in the electrodes and corresponds to electron-hole pair recombinations specified by $S_{LL}$, $S_{RR}$ and $S_{LR}$ \cite{Zamoum_PhysRevB.93.235449}. Nevertheless, by considering the coupling to the electric field only in the dielectric layer, we implicitly assume a relaxation in the tunneling barrier associated to the noise spectral density $S_{ii}=(S_{LL}+S_{RR}-2S_{LR})/4$ and our approach is not in contradiction with the inelastic interpretation. We actually use the elastic tunneling current to calculate the radiation impedance neglecting the feedback of the electromagnetic environment on the current fluctuations. This feedback, called the dynamical Coulomb blockade, is responsible for inelastic tunneling processes but is negligible here since $\mathcal{R}(\nu)\ll R_K$ \cite{Xu14,Altimiras16}.

\vspace{0.2cm}\noindent We have measured the current fluctuations $S_{ii}$ in a metallic tunnel junction in the optical domain. In this regime, $S_{ii}$ cannot be described anymore with a usual fluctuation dissipation relation because of the energy and voltage dependance of the tunneling transmission. We have shown how this dependence can be incorporated into the Landauer-B\"{u}ttiker formalism to ensure the gauge invariance of the $I(V)$ characteristic in the far-from-equilibrium regime and describe the quantum fluctuations of the current at optical frequencies. This theoretical description is in good agreement with our experimental results and sheds light on the estimation of quantum efficiency of metallic tunnel junction as a light emitter. Our experimental approach demonstrates that optical measurements are a powerful tool to study the quantum electronic transport at high energy ($\sim 1 \, \mathrm{eV}$) and extend the range of applicability of conventional concepts  of mesoscopic electronic transport. Establishing a new fluctuation dissipation relation in the optical regime will require a properly defined response function of the tunneling current at optical frequencies.

\vspace{0.2cm}\noindent\emph{Acknowledgements.} We acknowledge fruitful discussions with E. Akkermans, M. Aprili, J. Basset, E. Boer-Duchemin, J. Est\`{e}ve, J-J Greffet, B. Reulet, E. Pinsolle I. Safi and P. Simon. We also thank A. Cr\'{e}pieux for useful insight. This work was supported by ANR-11-JS04-006-01, Investissements d'Avenir LabEx PALM (ANR-10-LABX-0039-PALM) and  ANR-15-CE24-0020.

\section*{Supplemental Material}

This supplemental material provides details on (I) the experimental setup, (II) the calibration procedure used to extract the current shot noise in the zero frequency limit, (III) the Landau-B\"{u}ttiker formalism used to describe the current noise in the tunnel junction in the far-from-equilibrium regime (FFER), (IV) the validity of the fluctuation-dissipation relation (FDR) and (V) the Laks-Mills theory used to derive the radiation impedance.

\begin{figure}
\begin{center}
\includegraphics[width=1\linewidth]{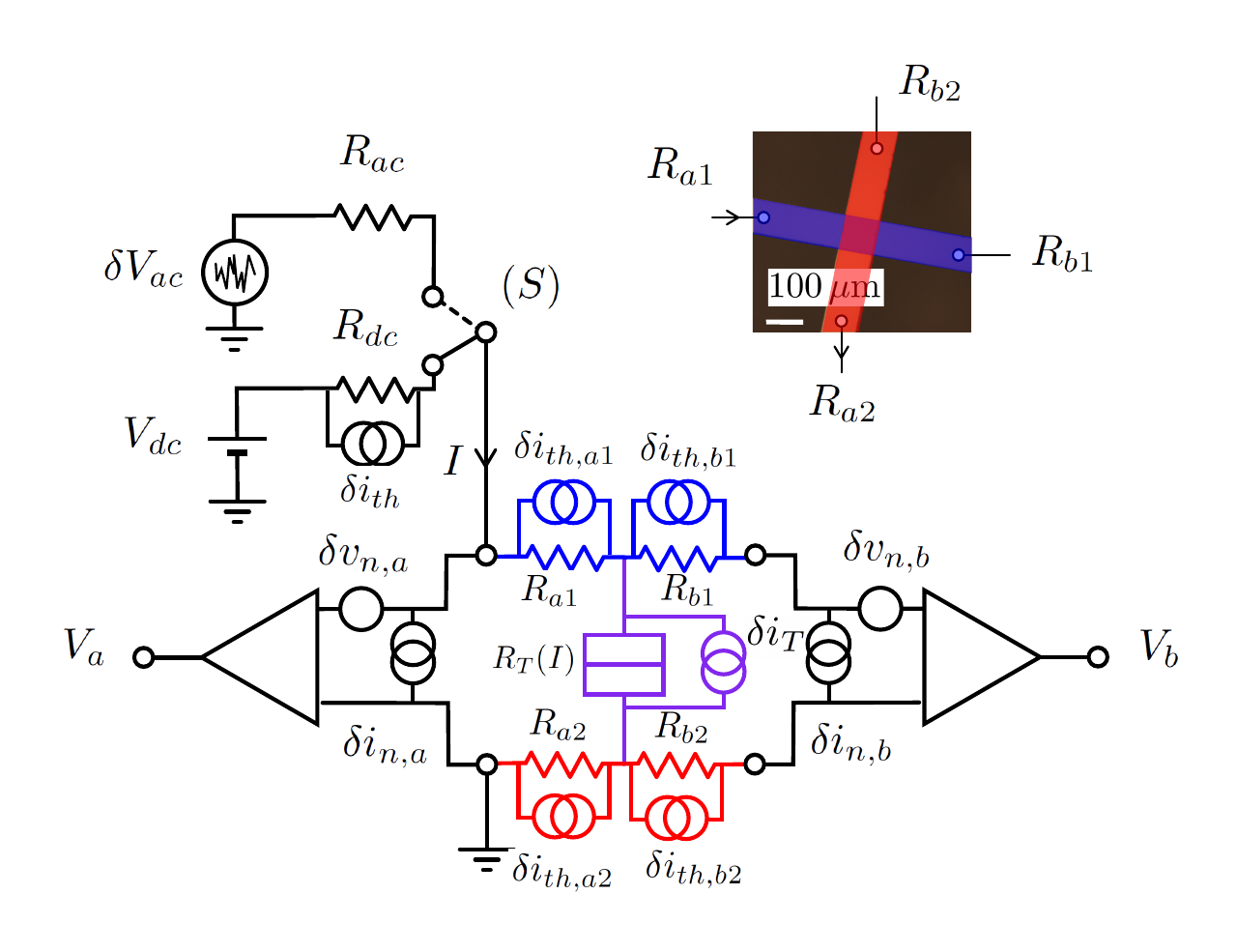}
\end{center}
\vspace{-0.5cm}
\caption{Detailed view of the experimental setup used for the current noise measurement in the zero frequency limit. \textit{Inset}:  optical micrograph of the metallic cross-junction including the resistive electrodes $R_{a1}$,$R_{a2}$,$R_{b1}$ and $R_{b2}$ .\label{FigS1}}
\end{figure}

\section{Sample fabrication and experimental setup}

The sample is a planar aluminum tunnel junction $SiO^{(20)}/Al^{(5)}/Al_20_3^{(2)}/Al^{(5)}/SiO^{(10)}$ deposited on a sapphire substrate. Numbers stand for the thickness in nm. The $100 \times 100 \, \mu \mathrm{m}^2$ junction is fabricated by thin film deposition through shadow masks in a typical base pressure of $10^{-9} \, \mathrm{mbar}$ with an oxidation of the first $Al$ electrode in an oxygen glow discharge. The Kretschmann configuration is realized by using a $BK7$ glass prism \cite{Kretschmann72}. The $I(V)$ characteristic is measured using a standard four points technique with a dc voltmeter whereas the bias-dependence of the tunneling conductance is measured using a standard lock-in technique. The current noise in the zero frequency limit is measured using a cross-correlation technique and a real time FFT-based spectral measurement performed with a digitizer.  Radiated power at wavelength $\lambda$  is measured with filtered Si (at $\lambda=0.9 \, \mu \mathrm{m}$) and InGaAs  (at $\lambda=1.3 \, \mu \mathrm{m}$) amplified detectors and a lock-in technique by modulating the voltage bias at $7 \, \mathrm{Hz}$.  Their noise equivalent power are $\sim 1.2 \times 10^{-14} \, \mathrm{W}/\sqrt{\mathrm{Hz}}$ and $\sim 6.0 \times 10^{-14} \, \mathrm{W}/\sqrt{\mathrm{Hz}}$ respectively. To collect as much light as possible, the emitted light is refracted on a conical prism then collimated on the photon detector by using an aspherical lens (focal length $f=8 \, \mathrm{mm}$, numerical aperture $NA=0.5$). The detection efficiency is estimated at $80 \%$.

\section{Calibration of the shot noise measurement setup}

\begin{figure}
\begin{center}
\includegraphics[width=0.8\linewidth]{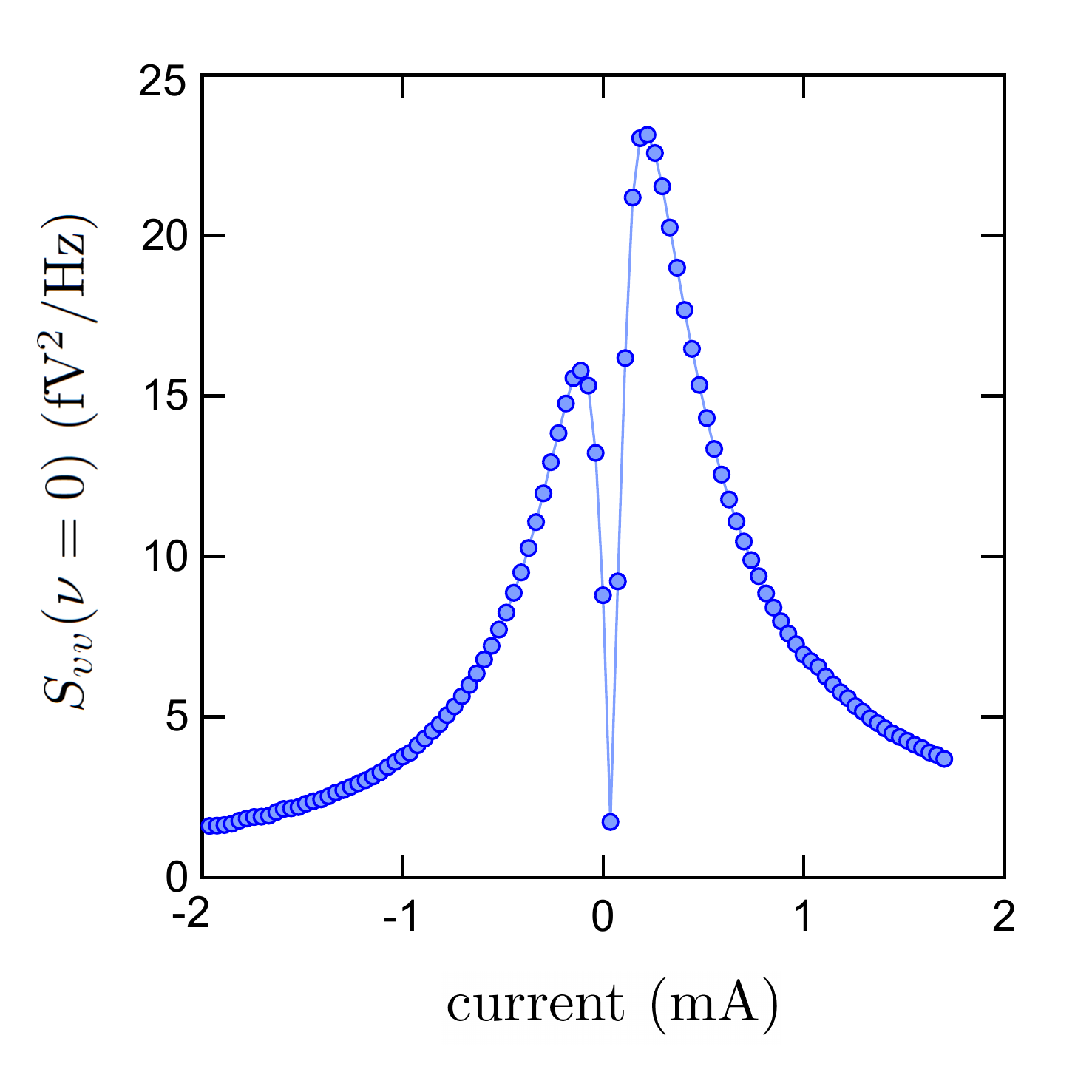}
\end{center}
\vspace{-0.5cm}
\caption{Voltage noise spectral density measured in the bandwidth $[10 \, \mathrm{kHz},100 \, \mathrm{kHz}]$.\label{FigS2}}
\end{figure}

\begin{figure}
\begin{center}
\includegraphics[width=0.8\linewidth]{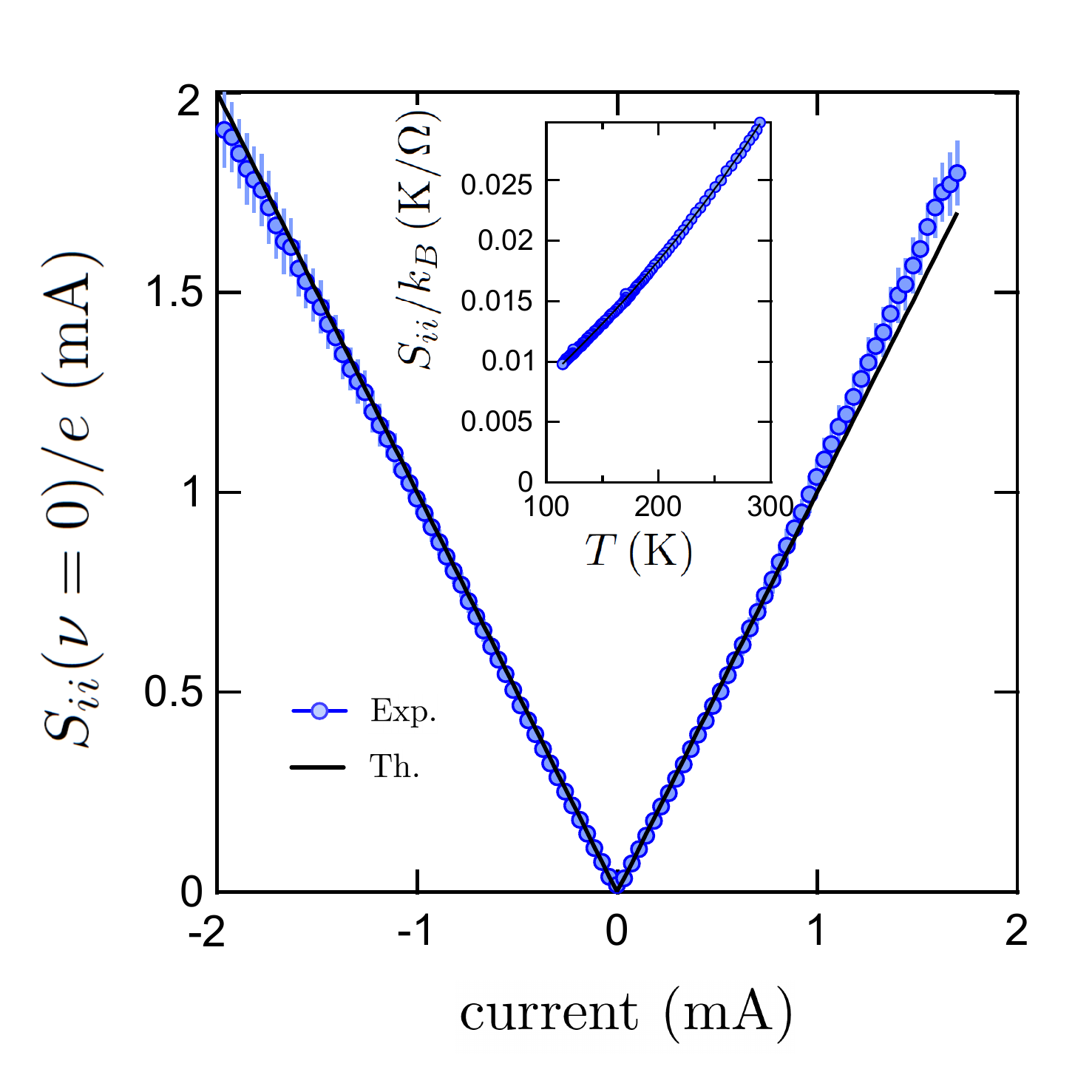}
\end{center}
\vspace{-0.5cm}
\caption{Current noise spectral density measured in the bandwidth $[10 \, \mathrm{kHz},100 \, \mathrm{kHz}]$. \textit{Inset}: thermal noise measured at zero voltage bias for different temperatures. Solid line corresponds to the fluctuation-dissipation theorem expectation.\label{FigS3}}
\end{figure}

The current noise in the zero frequency limit is measured using a cross-correlation technique to remove the amplifier voltage noise ($\delta v_n \simeq 2 \, \mathrm{nV}/\sqrt{\mathrm{Hz}}$). If the current noise ($\delta i_n \simeq 15 \, \mathrm{fA}/\sqrt{\mathrm{Hz}}$) can be neglected, the thermal noise of the contact resistances in series with the tunnel junction has to be subtracted. The resistance of the thin electrodes ($5 \, \mathrm{nm}$) are indeed of the same order of magnitude than the differential tunnel resistance $R_T \simeq 150 \, \Omega$ at high voltage bias: $R_{a1}=221 \, \Omega$, $R_{a2}=280 \, \Omega$, $R_{b1}=172 \, \Omega$, $R_{b2}=314 \, \Omega$ (see inset of FIG.~\ref{FigS1}). The voltage noise $S_{vv}(V)\equiv\langle V_a V_b\rangle$ measured by the experimental set up depicted in  FIG.~\ref{FigS1} is:
\small
\begin{equation}\label{eq.Svv}
S_{vv}(V)=g\left(|Z_{setup}(V)|^2 S_{ii}(V)+S_{vv,setup}(V)\right)
\end{equation}
\normalsize
\noindent where $g$ is the global gain of the amplifier chain, $Z_{setup}$  is the transimpedance and $S_{vv,setup}$ the excess noise related to the measurement setup.  FIG.~\ref{FigS2} shows the voltage noise spectral density $S_{vv}$ measured in the frequency range $[10 \, \mathrm{kHz},100 \, \mathrm{kHz}]$. It cannot be directly compared to the current noise spectral density $S_{ii}$ of the tunneling current because of the voltage dependence of $R_T(V)$. A white voltage noise source $\delta V_{ac}$ is then used to calibrate the detection setup. If $\delta V_{ac}$ is high enough to neglect the intrinsic noise of the junction, the measured voltage noise $S_{vv,cal}$ enables to determine $Z_{setup}$ and $S_{vv,setup}$:
\small
\begin{subequations}
\label{allequations}
\begin{align}
|Z_{setup}(V)|^2=&\frac{1+R_a/R_{dc}}{1+R_a/R_T(V)} \frac{S_{vv,cal}/g}{\left(\delta V_{ac}/R_{ac}\right)^2}\label{eq.Zsetup} \\
S_{vv,setup}(V)=&\frac{2k_BT_{300K}}{R_{dc}} \left(1+\frac{T/T_{300K}}{1+R_T(V)/R_a}\right) \nonumber\\
& \hspace{2cm} \times \frac{S_{vv,cal}/g}{\left(\delta V_{ac}/R_{ac}\right)^2}\label{eq.Svvesetup}\\
\nonumber
\end{align}
\end{subequations}
\normalsize
\noindent where $R_{dc}=10 \, \mathrm{k}\Omega$, $R_{ac}=10 \, \mathrm{M}\Omega$, $R_a=R_{a1}+R_{a2}\simeq 500 \, \Omega$, $T_{300K}=300 \, \mathrm{K}$ and $T$ is the temperature of electrons.  FIG.~\ref{FigS3} shows the current shot noise and the  theoretical expectation given by the FDR. The thermal noise measured at zero voltage bias for different temperature is shown on the inset of FIG.~\ref{FigS3} and is in good agreement with the fluctuation-dissipation theorem $S_{ii}(eV=0)=2k_BT/R_T$. Note the typical temperature dependence of the tunneling junction resistance which increases when the temperature decrease \cite{Patino15}. As mentioned in the article, the Joule heating ($P_J\sim 5 \, \mathrm{mW}$) cannot explain the discrepancy between the data and the theory. The electron-phonon coupling for $T\geq 100 \, \mathrm{K}$ gives a thermal conductance $G_{e-ph}\geq 10^{17} \mathrm{W}.\mathrm{m}^{-3}\mathrm{K}^{-1}$ which leads to an electronic temperature equals to the temperature $T_0$ of the lattice such as: $T-T_0 < 0.5 \, \mathrm{mK}$ \cite{Lin08}. One indeed deduces that the temperature of electrons is homogeneous over the whole sample.

\section{Electronic transport in a tunnel junction in the far from equilibrium regime}

\subsection{Energy and voltage dependence of the transmission - $I(V)$ characteristics}\label{section.param}

We consider a tunnel junction  with the surface area $S$ and a large number of transverse channels labeled by the wave vector $\mathbf{k}_{\perp}$. We assume that electrons are scattered elastically on the tunneling barrier without any inelastic energy loss inside the barrier. The tunnel current $I$ is then given by the Landauer-B\"{u}ttiker formula:

\small
\begin{equation}\label{eq.IV2}
I(V)=\frac{2e}{h} \, \sum_{k_{\perp}} \int_{0}^{+\infty} d\epsilon \,  \mathcal{T}_{3D}(\epsilon,\mathbf{k}_{\perp}) \left\{f(\epsilon-eV)-f(\epsilon)\right\}
\end{equation}
\normalsize

\noindent where $f(\epsilon)=[1+ \exp ((\epsilon-\epsilon_F)/k_BT)]^{-1}$ is the Fermi-Dirac distribution, $\epsilon_F$ the Fermi energy and $\mathcal{T}_{3D}(\epsilon,\mathbf{k}_{\perp})$ the transmission probability for an incoming electron with a transverse wave vector $\mathbf{k}_{\perp}$ and a total energy $\epsilon=\epsilon_{\perp}+\epsilon_{\parallel}$. To recover the expression of Eq.~(3) in the article, we define  the  transmission $\mathcal{T}(\epsilon,eV)$ the  transmission for an incoming  electron  with a total energy $\epsilon$ by averaging the WKB transmission over all possible values of $\epsilon_{\perp}$ \cite{Simmons64,Hansen04}:

\small
\begin{equation}\label{eq.T1D}
\mathcal{T}(\epsilon,eV)=\sum_{k_{\perp}}\mathcal{T}_{3D}(\epsilon,\mathbf{k}_{\perp}) = M \int_{-\infty}^{\epsilon} T_{WKB}(\epsilon_{\parallel},eV) \,\frac{d\epsilon_{\parallel}}{\epsilon_F}
\end{equation}
\normalsize

\noindent where $T_{WKB}(\epsilon_{\parallel},eV)$ is the Wentzel-Kramers-Brillouin (WKB) transmission coefficient through a 1D potential barrier $U(z)$, $d$ is the thickness of the barrier and $M=\pi S/\lambda_F^2$ is the number of transversal modes of conduction contained in the tunnel junction area $S$:

\small
\begin{equation}\label{eq.WKB}
T_{WKB}(\epsilon_{\parallel},eV) =\exp \left\{-\frac{\sqrt{8m}}{\hbar}\int_0^d \sqrt{U(z){+}eV\left(1{-}\frac{z}{d}\right){-}\epsilon_{\parallel}}\right\}dz
\notag
\end{equation}
\normalsize
\noindent We are considering here the total potential including the potential barrier $U(z)$ and the biasing energy $U_{bias}(z,V)=eV(1-z/d)$. The biasing energy considers only the energy of the tunneling electron in the uniform electric field induced by the bias voltage, we have implicitly neglected the effects of space charge inside the barrier and image charge in the electrodes. In aluminum, the Fermi energy is $\epsilon_f=11.7 \, \mathrm{eV}$ and the Fermi wavelength $\lambda_F=0.36 \, \mathrm{nm}$. Then, the number of channels in the tunnel junction  is  $M \sim 2.4 \times 10^{11}$ and the transmission in the considered voltage range is $\mathcal{T}<1.7 \times 10^{-11}$. The asymmetry of the trapezoidal barrier is obtained with the second order expansion of the normalized conductance \cite{Brinkman1970}:
\small
\begin{equation}\label{eq.G0}
\frac{G(V)}{G(0)}=1+\frac{V}{V_1}+\left(\frac{V}{V_2}\right)^2,
\end{equation}
\normalsize

\noindent with $\Delta U/U=3\sqrt{2}V_2/V_1$. The parabolic fit of data in the inset of Fig. \ref{FigS2} gives $V_1 \simeq -2.85 \pm 1 \, \mathrm{V}$ and $V_2 \simeq 0.78 \pm 0.06 \, \mathrm{V}$. We then deduce $\Delta U/U \simeq 1.1$. The values of $U \simeq 2.68 \, \mathrm{eV}$ and $d \simeq 2 \, \mathrm{nm}$ are estimated from the fit of the $I(V)$ characteristics. These values are obtained by considering the effective mass of electrons in the oxide ($m=0.38 \times 9.1 \times 10^{-31} \, \mathrm{kg}$) \cite{Zemanova_PhysRevB.87.195107}. We have checked that the charging effects in the barrier slightly change $U$ and $d$ of about $10 \%$. The large value of the asymmetry $\Delta U/U$ can be attributed to the growth on different substrates ($SiO_2$/$AlO_x$). One has to keep in mind that the trapezoidal barrier model is a simplistic model which cannot fully describe our sample. The effects of image charge could be considered in the potential barrier $U$, they would only re-normalized the barrier height. They will be taken into account only to estimate the tunneling current. The capacitance  $\sim 0.5\, \mathrm{nF}$  of the tunnel junction is measured thanks to the cut-off frequency observed on the noise spectral density at low bias voltage. This value is in agreement with the thickness of the tunnel barrier: $C= \epsilon_r \epsilon_0 S/d \simeq 0.43 \, \mathrm{nF}$ where $\epsilon_r=9.8$ is the dielectric constant of alumina, $\epsilon _0$ is the vacuum permittivity and $S$ the surface of the junction.

\subsection{Gauge invariance}

The gauge transformation corresponds to the  addition of a constant potential $V_0$ on both electrodes. It leads to the following transformations: $eV(1-x/d) \rightarrow  eV(1-z/d)+eV_0$ for the biasing energy, $T_{WKB}(\epsilon,eV) \rightarrow T_{WKB}(\epsilon-eV_0,eV)$ for the WKB transmission coefficient and $f(\epsilon-eV) \rightarrow f(\epsilon-e(V+V_0))$ and $f(\epsilon) \rightarrow f(\epsilon-eV_0)$ for the Fermi-Dirac distributions in the electrodes. It is straightforward to check that Eq~(3)(4a)-(4c) in the article are invariant under these transformations.

\subsection{Current noise spectral density at finite frequency}

\begin{figure}
\begin{center}
\includegraphics[width=0.8\linewidth]{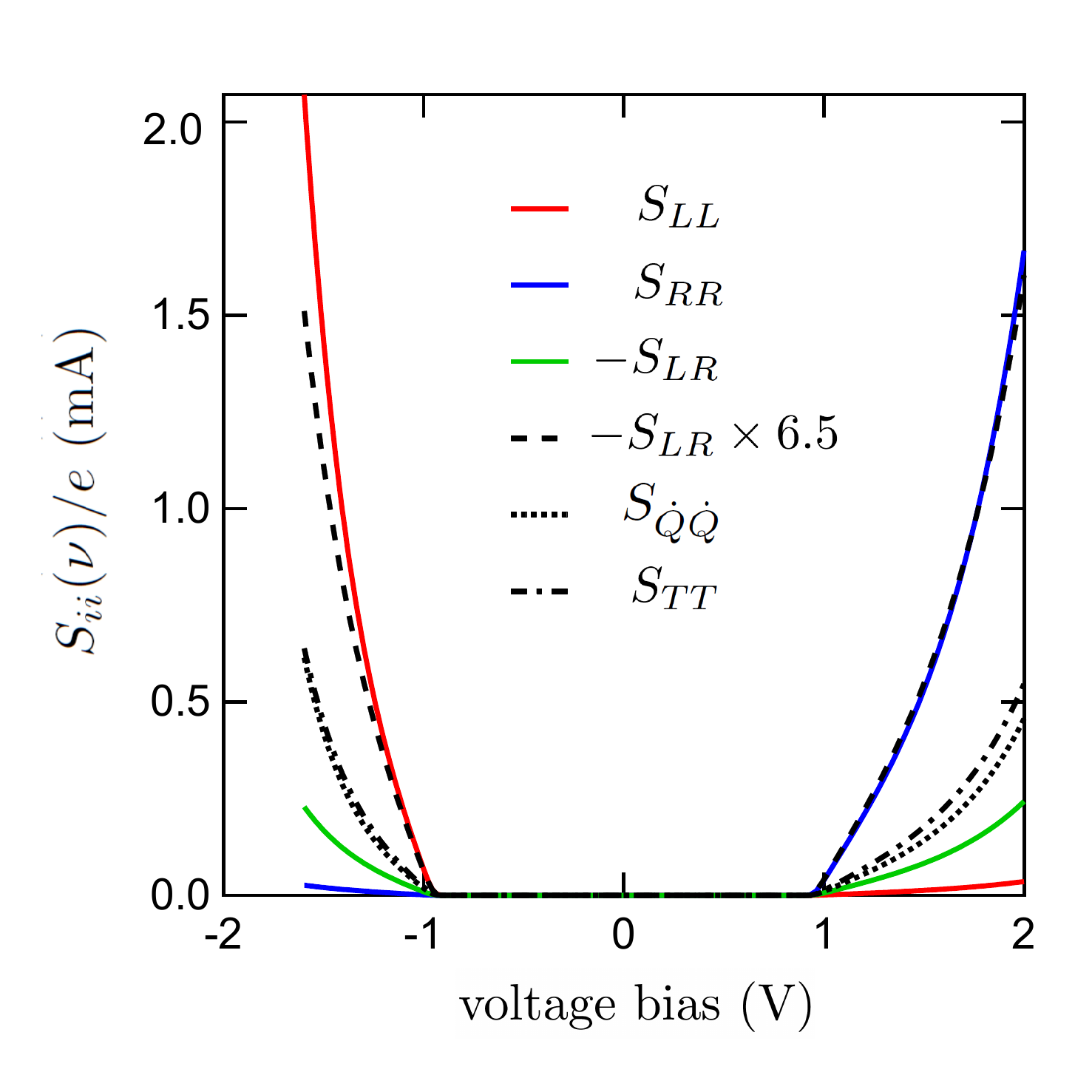}
\end{center}
\vspace{-0.5cm}
\caption{Theoretical current spectral noise density $S_{LL}$, $S_{RR}$ and $S_{LR}$ in the high frequency limit ($\lambda=1.3 \, \mu \mathrm{m}$) for the trapezoidal barrier used to fit the experimental $I(V)$ characteristics.\label{FigS4}}
\end{figure}

At zero frequency, the current noise spectral density is given by the fluctuation dissipation theorem  $S_{ii}(h\nu)=2G \, h\nu \, N(h\nu)$ where $N(\epsilon)=1/((\mathrm{exp}(\epsilon/k_BT)-1)$ denotes the Bose-Einstein distribution and $G$ the dc conductance  of the junction. Note that the absorption is due to the tunnel resistance and not to the resistance of the electrodes which are assumed negligible compared to the resistance of junction. However, as it has been shown in the article, the noise spectral density depends on the electrode where it is evaluated because of the energy and voltage dependent transmission ($S_{LL}\neq S_{RR}\neq -S_{LR}$). It should also be stressed that, if we only consider the energy dependence of the transmission and omit its voltage dependence, the gauge invariance is  violated and only one of the correlators satisfies the FDR, $S_{LL}=S_{ii}^{(FDR)}$ according to our choice of voltage biasing. For a 3D tunnel junction, Eq.~(4c) in the article has to be slightly modified to take into account the summation over the transversal modes:
\small
\begin{eqnarray}\label{eq.SLR3D}
S_{LR}(eV,h\nu)=-\frac{Me^2}{h}\int d\epsilon \int_{-\infty}^{\epsilon}\frac{d\epsilon'}{\epsilon_F} \sqrt{\mathcal{T}_{WKB}(\epsilon',eV)}\nonumber\\
\times \sqrt{\mathcal{T}_{WKB}(\epsilon'{-}h\nu,eV)}\left\{ f_L(\epsilon)(1{-}f_R(\epsilon{-}h\nu)) \right.\nonumber\\
 \left.+ f_R(\epsilon)(1{-}f_L(\epsilon{-}h\nu))\right\}
\end{eqnarray}
\normalsize
\noindent whereas Eqs.~(4a) and (4b) remain unchanged considering the transmission given by Eq.~(\ref{eq.T1D}). FIG.~\ref{FigS4} shows the theoretical noise spectral density  $S_{LL}$, $S_{RR}$ and $S_{LR}$ at $\lambda=hc/\nu=1.3 \, \mu \mathrm{m}$ using the parameters $(U,\Delta U,d)$ of the junction. $S_{LL}$ and $S_{RR}$ exhibit a strong dissymmetry revealing that energy relaxation occurs essentially in the left (\textit{resp.} right) electrode for $eV<0$  (\textit{resp.} $eV>0$) and can be interpreted as an electron-hole pairs recombination in the left  (\textit{resp.} right) electrode \cite{Zamoum_PhysRevB.93.235449}. $S_{LR}$ is more difficult to interpret and appears as an interference between the two former processes. Note that $S_{LR}$ is almost proportional to $S_{LL}$ (\textit{resp.} $S_{RR}$) for $eV>0$ (\textit{resp.} $eV>0$). It implies, because of the strong asymmetry, $S_{LR} \simeq -\sigma(\nu) \left( S_{LL}+S_{RR}\right)$ with $\sigma>0$ a factor of proportionality depending on the frequency $\nu$ (see Fig. \ref{FigS4}). The tunneling current is assumed to be constant in the barrier and given by the average current $I_T=(I_L-I_R)/2$. The tunneling current spectral noise density is then:
\small
\begin{equation}\label{eq.ST}
S_{TT}(eV,h\nu)=\frac{1}{4} \left(S_{LL}+S_{RR}-2S_{LR} \right)
\end{equation}
\normalsize
\noindent Although photon emission is due to the coupling to the fluctuations of the tunneling current $S_{TT}$, we can compare this quantity to the fluctuations of the accumulated charges on the electrodes of the junction related to $dQ/dt=(I_L+I_R)/2$:
\small
\begin{equation}\label{eq.SQ}
S_{\dot{Q}\dot{Q}}(eV,h\nu)=\frac{1}{4} \left(S_{LL}+S_{RR}+2S_{LR} \right)
\end{equation}
\normalsize
\noindent FIG. \ref{FigS4} and FIG. \ref{FigS5} show a significant difference between $S_{TT}$ and $S_{\dot{Q}\dot{Q}}$. We also notice that $S_{\dot{Q}\dot{Q}} = 0$ at zero frequency. The experimental data presented in the article in Fig. 4 falls on $S_{TT}$ and are not in agreement with $S_{\dot{Q}\dot{Q}}$.

\subsection{Traversal time in a tunnel junction}

The time for an electron to cross the barrier is defined as the traversal time $\tau=d\sqrt{m/(2(U-eV))}$ where $U$ is the barrier height, $d$ its thickness, $m$ the effective mass of electron and $V$ the bias voltage. For a common aluminum oxide barrier $U \sim 2 \, \mathrm{eV}$, $d \sim 1 \, \mathrm{nm}$ and $m \sim 3.5 \times 10^{-31} \, \mathrm{kg}$ which gives $\tau \sim 10^{-15} \,\mathrm{s}$ at $1 \, \mathrm{V}$ \cite{Buttiker_PhysRevLett.49.1739}. This time is comparable to the time scale probed by the spectral noise density at optical frequencies. It is also comparable to the average time $\tau_Q=h/eV$ between electrons emitted between the two voltage biased electrodes which gives $\tau_Q \simeq 4 \times 10^{-15} \, \mathrm{s}$ at $1 \, \mathrm{V}$  \cite{Gabelli_15}.

\section{Fluctuation-dissipation relation at zero frequency}
We give here a derivation of the FDR at zero frequency using the steady state fluctuation theorem (SSFT). This theorem results in a generalization of the second law of thermodynamics and holds under very general hypothesis \cite{PhysRevLett.71.2401,PhysRevB.75.155316}. We describe the electronic transport through the tunnel junction as a charge transfer where $\Gamma_{+/-}$ stands for the probability per unit time to transfer an electron from the left/right electrode to the right/left electrode. Note that no particular hypothesis is made on the transfer rates $\Gamma_{+/-}$. The resulting probability $p(q)$ to transfer a charge $q$ during a tunneling event is given by:
\small
\begin{equation}\label{eq.proba1}
\begin{split}
p(q)=&\left(1-(\Gamma_{+}+\Gamma_{-}) \delta t\right)\, \delta(q) \\
 & \hspace{0.5cm}  +\Gamma_{+} \delta t\, \delta(q-e)+\Gamma_{-} \delta t\, \delta(q+e)
\end{split}
\end{equation}
\normalsize
\noindent where $\delta t$ is the characteristic time of the tunneling event. In the long time limit ($\Delta t = N\delta t \rightarrow + \infty$), the charge $Q$ transferred through the junction is the sum of $N$ independent random variables $Q=\sum_{i=1}^{N}q_i$ and its distribution probability reads:
\small
\begin{equation}\label{eq.proba2}
P(Q)=\int \prod_{i=1}^{N} dq_i \, p(q_i) \delta \left(Q-\sum_{i=1}^{N}q_i\right)
\end{equation}
\normalsize
\noindent Let's introduce the moment generating function $\chi_Q(\lambda)$ which offers a convenient way to characterize the distribution function $P$:
\small
\begin{equation}\label{eq.chi}
\begin{split}
\chi_Q(\lambda)&=\int dQ \, P(Q)e^{i\lambda Q} \\
 & = \left( (1- (\Gamma_{+}+\Gamma_{-})\delta t) +\Gamma_{+} \delta t e^{i\lambda e}+\Gamma_{-} \delta t e^{-i\lambda e} \right)^N
\end{split}
\end{equation}
\normalsize
\noindent which becomes in the tunneling limit ($\Gamma_{+/-} \delta t \ll 1$):
\small
\begin{equation}\label{eq.chitunnel}
\chi_Q(\lambda) \simeq  1- N\delta t \left( (\Gamma_{+}+\Gamma_{-}) -\Gamma_{+} \delta t e^{i\lambda e}-\Gamma_{-} \delta t e^{-i\lambda e} \right)
\end{equation}
\normalsize By applying the SSFT to a voltage bias tunnel junction, $P(Q)/P(-Q)=e^{-QV/k_BT}$, we get:
\small
\begin{equation}\label{eq.SSFT}
\chi_Q\left(\lambda\right)=\chi_Q\left(-\lambda -i \frac{V}{k_BT}\right)
\end{equation}
\normalsize allowing to deduce a detailed balance relation between the transfer rate coefficients $\Gamma_{+/-}$:
\small
\begin{equation}\label{eq.balance}
\frac{\Gamma_{+}}{\Gamma_{-}}=\exp \left(- \frac{eV}{k_BT} \right)
\end{equation}
\normalsize The current $\langle I \rangle=\langle Q \rangle /\Delta t$ and the current fluctuations $\langle \Delta I^2 \rangle=\langle  \Delta Q^2 \rangle /\Delta t^2$ are then given by the first two terms of the Taylor expansion of the generating function:
\small
\begin{subequations}
\label{allequations}
\begin{eqnarray}
\langle I \rangle&=&e \left(\Gamma_{+}-\Gamma_{-}\right)\\
\langle \Delta I^2 \rangle&=&e^2 \left(\Gamma_{+}+\Gamma_{-}\right)\Delta f
\end{eqnarray}
\end{subequations}
\normalsize
where $\Delta f=1/\Delta t \rightarrow 0$ is the frequency bandwidth of the measurement. We finally obtain the FDR:
\small
\begin{equation}\label{eq.Sii0}
S_{ii}(eV,h\nu=0)= \frac{\langle \Delta I^2 \rangle}{\Delta f} = \frac{eI(V)}{\tanh \left( eV/2k_BT \right)}
\end{equation}
\normalsize
\section{Validity of the fluctuation-dissipation relation at finite frequency}
\begin{figure}
\begin{center}
\includegraphics[width=1\linewidth]{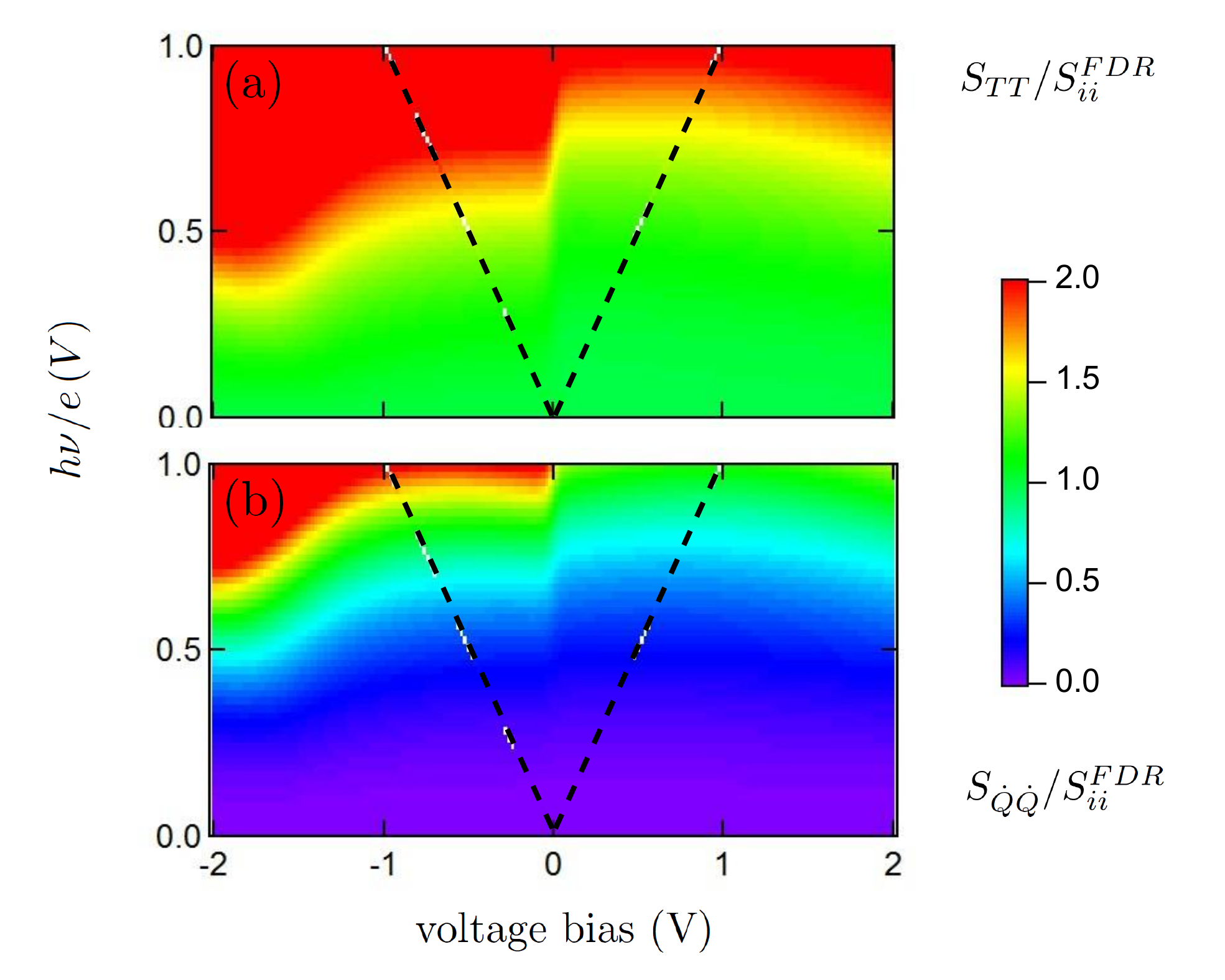}
\end{center}
\vspace{-0.5cm}
\caption{(a) Color plot of the ratios $S_{TT}/S_{ii}^{(FDR)}=(S_{LL}+S_{RR}-2S_{LR})/(4S_{ii}^{(FDR)})$. (a) Color plot of the ratios $S_{\dot{Q}\dot{Q}}/S_{ii}^{(FDR)}=(S_{LL}+S_{RR}+2S_{LR})/(4S_{ii}^{(FDR)})$. Dashed lines correspond to the cross-over $eV=\pm h\nu$\label{FigS5}}
\end{figure}

As it has been shown in the article, $S_{ii}\neq S_{ii}^{(FDR)}$ as soon as $h\nu \neq 0$. However, the ratio $S_{ii}/S_{ii}^{(FDR)}$ is nearly voltage independent for $h\nu<1 \, \mathrm{eV}$ (see FIG.~\ref{FigS5}). It means that, even if $S_{ii}^{(FDR)}$ could approximatively  explain the voltage dependence of the emitted light power $P_{\nu}$, the radiation impedance would be overestimated because $S_{ii}$ is underestimated. In reference \cite{Roussel_PhysRevB.93.045102}, Roussel \textit{et al.} show the validity of the FDR provided few hypothesis. They use the non-equilibrium Kubo formula \cite{Kubo57,Safi10},
\small
\begin{equation}\label{eq.Kubo}
S_{ii}(eV,-h\nu)-S_{ii}(eV,h\nu)=2 h\nu \mathrm{Re}\left(G(eV,h\nu)\right),
\end{equation}
\normalsize
\noindent combined with the photon-assisted tunneling formula,
\small
\begin{equation}\label{eq.Gph}
\mathrm{Re}\left(G(eV,h\nu)\right)=e\frac{I(eV+h\nu)-I(eV-h\nu)}{2h\nu},
\end{equation}
\normalsize
\noindent where $G(eV,h\nu)$ is the non-equilibrium ac conductance measured at frequency $\nu$ for a dc voltage bias $V$.  However, Eq.~(\ref{eq.Gph}) does not hold for a voltage dependent transmission which is responsible for the FDR violation \cite{Gavish02}. It is also important to notice that $G(eV,h\nu)$ is not well defined at optical frequencies because of transversal dependence of the ac voltage related to the SPP excitation on the electrode of the tunnel junction. We also may ask questions about the validity of the LB approach at optical frequencies. We only use it to calculate the tunneling current which couples to the electric field in the barrier.
The LB formalism assumes that the electron wave vector is constant, equals to the Fermi wave vector $k_F$. This assumption is valid since we are considering electrons with energy $\epsilon$ close to the Fermi energy ($\epsilon _F = 11.7 \, \mathrm{eV}$ in aluminum). At optical frequencies $\nu$, $|\epsilon -\epsilon_F| \sim h\nu \sim 1 \, \mathrm{eV}$ and the current becomes position dependent on a typical length scale $l \sim \frac{2\epsilon_F}{h\nu}\lambda_F  \sim 7 \, \mathrm{nm}$ which remains larger than the electrode thickness.

\section{Surface plasmon polariton modes in the tunnel junction - Radiation impedance}

\begin{figure}
\begin{center}
\includegraphics[width=0.8\linewidth]{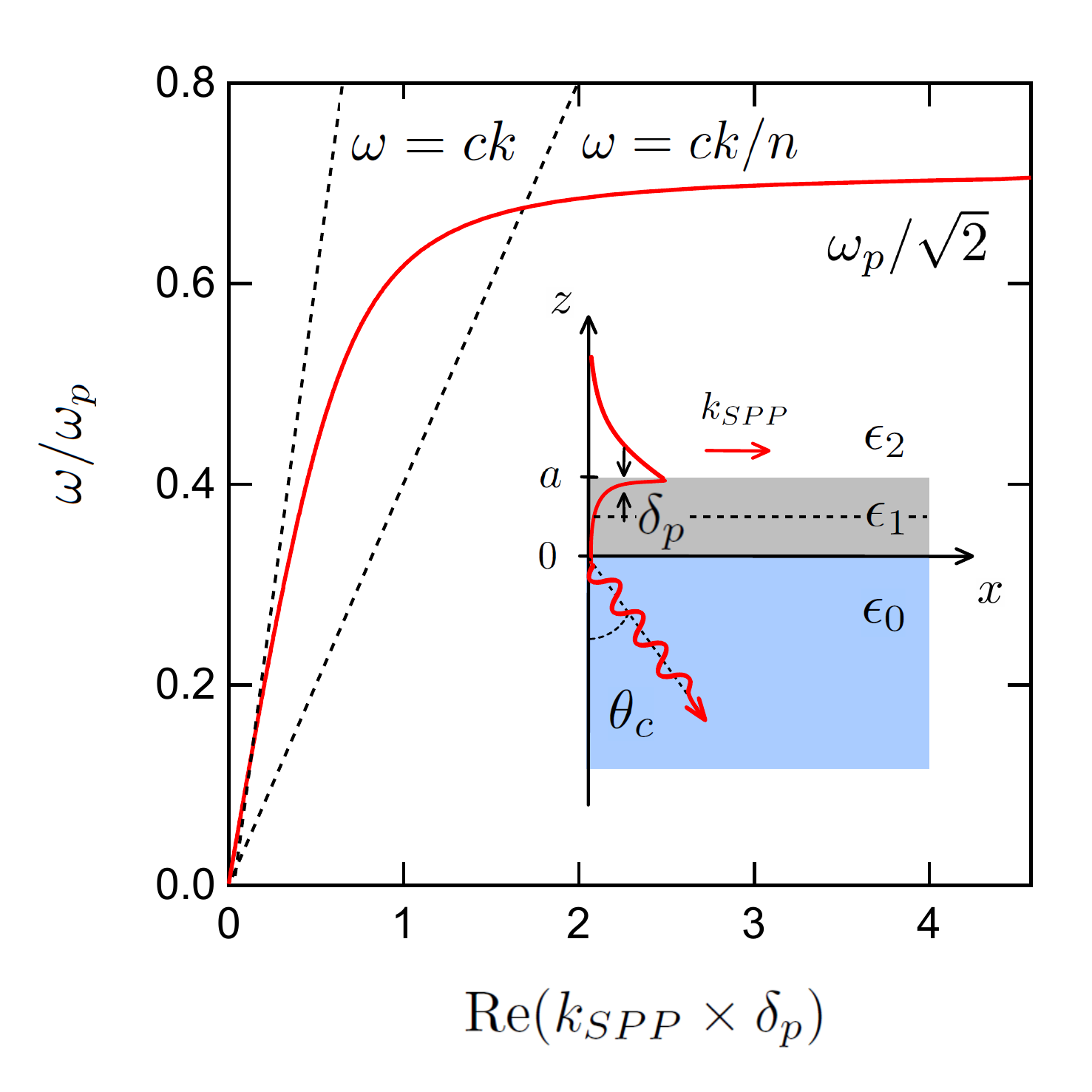}
\end{center}
\vspace{-0.5cm}
\caption{Dispersion relation of the SPP mode at the vacuum interface. Dashed lines correspond to dispersion relations in vacuum ($k=\omega/c$) and sapphire ($k=n\omega/c$) respectively. \textit{Inset}: geometry of the tunnel junction in contact with its sapphire substrate. The Kretschmann geometry enables the leakage of the SPP in the substrate at specific angle $\theta_p$. Dashed line corresponds to the tunneling barrier.\label{FigS6}}
\end{figure}

We consider here a simple tunnel junction made of two thick metallic layers with a total thickness $a$ separated by a thin layer of insulator. We can therefore distinguish two kinds of surface plasmon polariton (SPP) modes, the fast modes localized at the surfaces of the electrodes and the slow mode localized inside the tunneling barrier of thickness $d \ll a$. However, only the fast mode at the vacuum interface is coupled to the propagating mode in the sapphire substrate (see inset in FIG.~\ref{FigS6}). In the following, we then model the junction by a single metallic film of thickness $a$. In the preliminary approximation, we can consider the dispersion relation of a semi-infinite metallic layer \cite{Maier07}:
\small
\begin{equation}\label{eq.kSPP0}
k_{SPP,\infty}=\frac{\omega}{c} \sqrt{\frac{\epsilon_1\epsilon_2}{\epsilon_1+\epsilon_2}}
\end{equation}
\normalsize
\noindent where $\epsilon_1=1-(\omega_p/\omega)^2-i\gamma_p \omega_p^2/\omega^3$ is the Drude dielectric constant of the metal described by the plasma frequency $\omega_p$ and the damping term $\gamma_p$  and $\epsilon_2=1$ is the dielectric constant of the vacuum. In aluminum, reference \cite{Ordal85} gives $\omega_p=14.7 \, \mathrm{eV}$, $\gamma_p=80 \, \mathrm{meV}$ and $\delta_p=12.7 \, \mathrm{nm}$. FIG.~\ref{FigS6} shows the theoretical expectation of Eq.~(\ref{eq.kSPP0}). In the low frequency limit $\omega\ll\omega_p$, the fast SPP mode reduces to $k_{SPP}=\omega/c+\delta k_{SPP}$ with $\delta k_{SPP}\ll \omega/c$ and can leak in the substrate at the specific angle $\theta_p \simeq \arcsin 1/n$ such that $k_{SPP}=n \frac{\omega}{c} \sin \theta_p$.

\subsection{Coupling in the Kretschmann configuration}

\begin{figure}
\begin{center}
\includegraphics[width=0.8\linewidth]{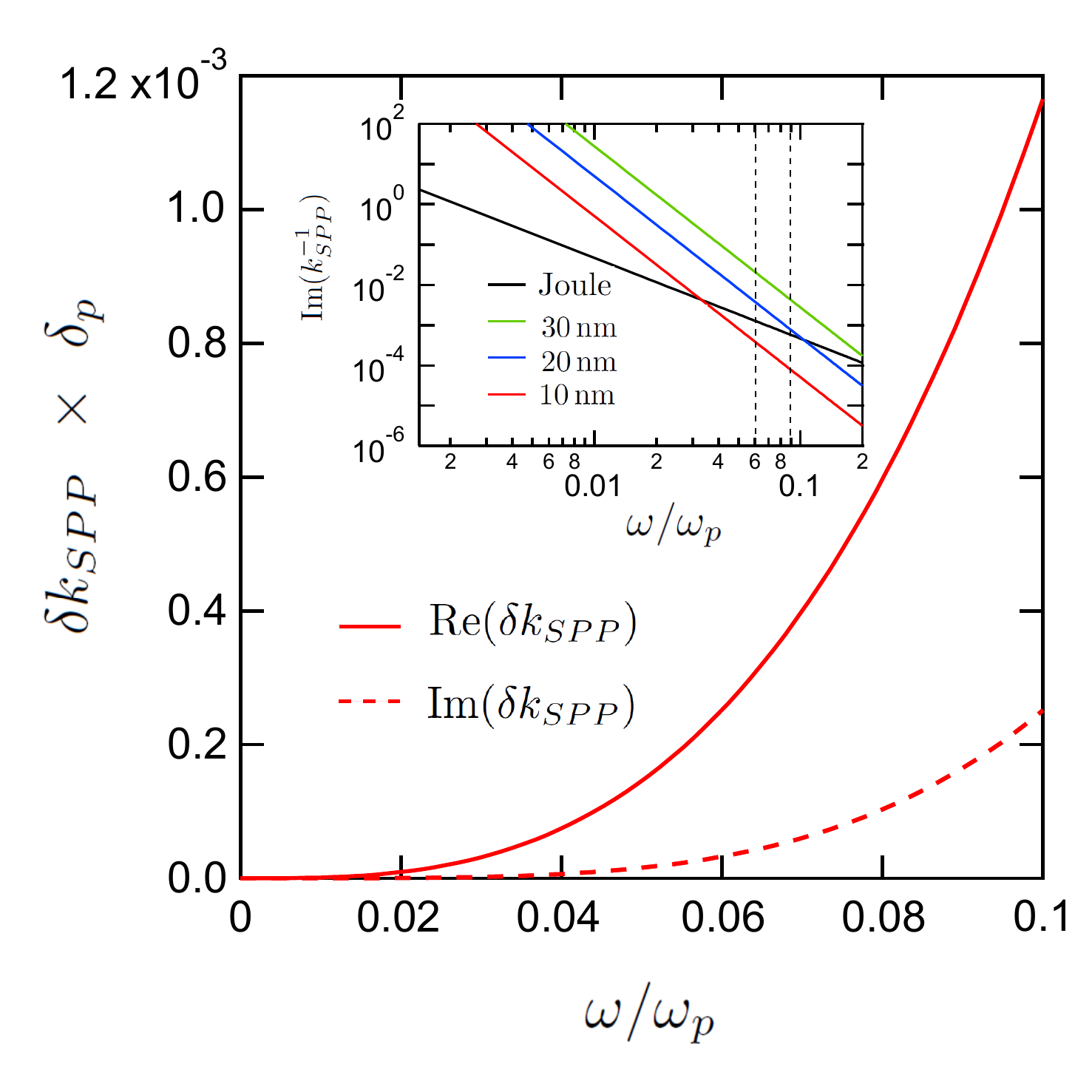}
\end{center}
\vspace{-0.5cm}
\caption{Dispersion relation $\delta k_{SPP}=k_{SPP}-\omega/c$ in the low frequency limit $\omega \ll \omega_p$. \textit{Inset}: coupling length $\mathrm{Im}(\delta k_{SPP}^{-1})$ in the Kretschmann configuration.\label{FigS7}}
\end{figure}

We now consider a thin metallic layer of thickness $a$ deposited on a substrate characterized by a dielectric constant $\epsilon_0=n^2$ to evaluate the leakage radiation (see inset in FIG.~\ref{FigS6}). We can assume that the thickness $d \ll a$ of the tunneling barrier has no effect on the field in the metallic electrodes. The $z$ component of the electric field in the junction is expressed by:
\small
\begin{subequations}
\label{allequations}
\begin{align}
E_0^{+} \exp(ik_{z0}z)+E_0^{-} \exp(-ik_{z0}z)& \hspace{0.2cm} \mathrm{for}\hspace{0.5cm} z<0\label{eq.Ez0}\\
E_1^{+} \exp(ik_{z1}z)+E_1^{-} \exp(-ik_{z1}z)& \hspace{0.2cm} \mathrm{for}\hspace{0.5cm} 0<z<a\label{eq.Ez1} \\
E_2^{+} \exp(ik_{z2}z) \hspace{2.7cm} & \hspace{0.2cm} \mathrm{for}\hspace{0.5cm} a<z\label{eq.Ez2} \\
\nonumber
\end{align}
\end{subequations}
\normalsize
\noindent with $\mathrm{Im}(k_{z,i})>0$. At the lowest non-trivial order in $\delta k_{SPP}$:
\small
\begin{subequations}
\label{allequations}
\begin{align}
k_{z0}\simeq&\frac{\omega}{c} \sqrt{n^2-1}\label{eq.kz0}\\
k_{z1}\simeq& \, i\frac{\omega_p}{c}\label{eq.kz1} \\
k_{z2}\simeq&\sqrt{-2\delta k_{SPP}\frac{\omega}{c}}\label{eq.kz2} \\
\nonumber
\end{align}
\end{subequations}
\normalsize
\noindent By implementing the boundary conditions of continuity of the electric and magnetic field parallel to the surface, we get:
\small
\begin{equation}\label{eq.E0p}
\frac{E_0^+}{E_2^+}= \frac{1}{2\sqrt{1-\beta^2}} \left\{\left(\frac{\epsilon_2}{\epsilon_0}+\frac{k_{z2}}{k_{z0}}\right) + \beta \left(\frac{\epsilon_1}{\epsilon_0}\frac{k_{z2}}{k_{z1}}+\frac{k_{z1}}{k_{z0}}\frac{\epsilon_2}{\epsilon_1}\right) \right\}
\end{equation}
\normalsize
\noindent with $\beta=-\tanh(ik_{z1}a)$. The dispersion relation $k_{SPP}(\omega)=\omega/c+\delta k_{SPP}$ is then solution of equation $E_0^+(k_{SPP})=0$ which gives at the first non-trivial order:
\small
\begin{equation}\label{eq.dkSPP}
\delta k_{SPP}\simeq\frac{1}{2\delta_p} \left(\frac{\omega}{\omega_p}\right)^3 \left\{1-2i \frac{\beta^2-1}{\beta}\frac{n^2}{\sqrt{n^2-1}} \left(\frac{\omega}{\omega_p} \right)\right\}
\end{equation}
\normalsize
\noindent FIG.~\ref{FigS7} shows $\delta k_{SPP}$ as a function of frequency in the low frequency limit $\omega \ll \omega_p$. Its inset compares the coupling length
$\mathrm{Im}(\delta k_{SPP}^{-1})$  for different thickness to the Joule dissipation length:
\small
\begin{equation}\label{eq.imk_joule}
\mathrm{Im}(\delta k_{SPP,\infty}^{-1}) \simeq \frac{1}{2} \frac{\gamma_p}{c} \left(\frac{\omega}{\omega_p}\right)^2
\end{equation}
\normalsize
\noindent It confirms that radiative damping is dominating for our experimental parameters $a=10 \, \mathrm{nm}$ and $0.064<\omega/\omega_p<0.1$.

\subsection{Electric and magnetic fields in the tunnel junction}

\begin{figure}
\begin{center}
\includegraphics[width=0.8\linewidth]{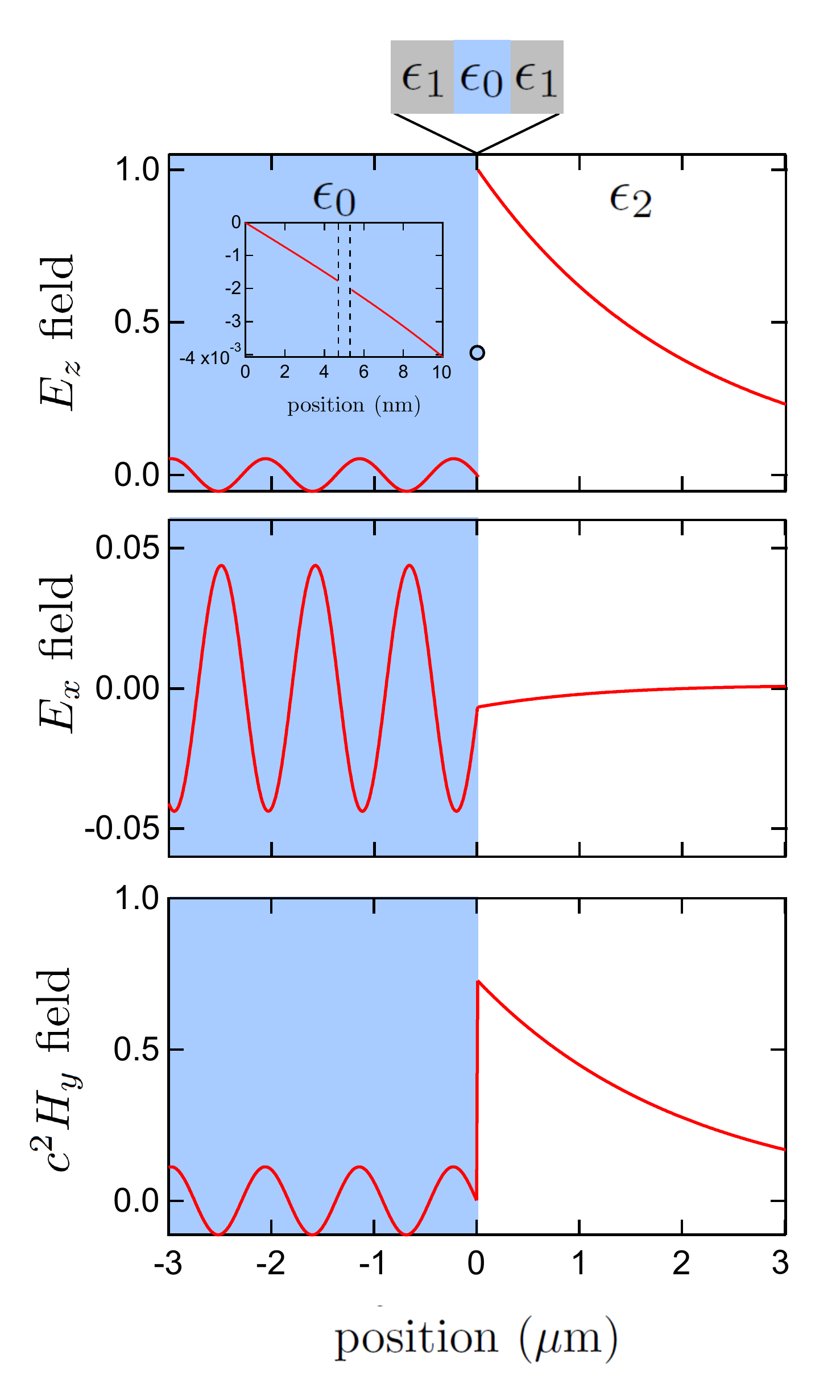}
\end{center}
\vspace{-0.5cm}
\caption{Electric $(E_x,0,E_z)$ and magnetic $(0,H_y,0)$ fields of the $\lambda=1.3 \, \mu \mathrm{m}$ SPP fast mode. They are normalized with respect to the field $E_z$ at position $z=d$. The blue circle corresponds to the value of the field in the thin ($d\sim 2\, \mathrm{nm}$) $Al_2O_3$ layer which corresponds to the tunnel barrier. \textit{Inset of the upper graph}: $z-$component of the electric field in the electrodes.\label{FigS8}}
\end{figure}

FIG~\ref{FigS8} shows the profile of the electric $(E_x,0,E_z)$ and magnetic $(0,H_y,0)$ fields components at $\omega/\omega_p=0.064$. $E_x$ and $H_y$ are continuous whereas $E_z$ exhibits discontinuities. The $z-$component of the electric field inside the tunnel barrier can be considered constant and is given at the third order in $\beta$ by:
\small
\begin{equation}\label{eq.EzT}
\frac{E_T}{E_2^{+}}=\frac{\epsilon_1}{\epsilon_0}E_{z1}\left(\frac{a}{2}\right) = \frac{1}{\sqrt{2}n^2}\left(1-\frac{\beta^2}{8}+ O(\beta^4)\right)
\end{equation}
\normalsize
\noindent where $\epsilon_0=n^2$ is also the dielectric constant of the alumina $Al_2O_3$ which is the same as sapphire. Note that the mode in the substrate is oscillating due to the radiative leakage of the SPP in the Kretschmann configuration.

\subsection{Radiation impedance in the low frequency limit $\omega \ll \omega_p$}

The Laks-Mills theory of light emission in a tunnel junction gives a radiated spectral power as a function of the two point spectral noise density $S_{ii}(z,z')$ and the $z-$component of the electric field inside the tunneling barrier \cite{Laks79,Laks80,Ushioda86}:
\small
\begin{eqnarray}\label{eq.Pnu}
P_{\nu}= Z_{vac}\int_{0}^{2\pi}d\varphi\int_{0}^{\pi/2}d\theta \, \frac{\omega^2 \sin ^3 \theta}{4\pi^2c^2}\int_{0}^{d}dz\,dz'\nonumber\\
\times \frac{E_{z1}(z)E_{z1}^{\star}(z')}{\left|E_0^+(k_x=n \frac{\omega}{c} \sin \theta)\right|^2} S_{ii}(z,z')
\end{eqnarray}
\normalsize
\noindent where $Z_{vac}=\sqrt{\mu_0/\epsilon_0} \simeq 376.7 \, \Omega$ is the vacuum impedance. By assuming a position independent electric field ($d\ll \delta_p$) and a position independent tunneling current ($I_T=(I_L-I_R)/2$), we get the radiation impedance:
\small
\begin{equation}\label{eq.RadImp}
\mathcal{R}=Z_{vac} \int_{0}^{\pi/2}d\theta \, \frac{\omega^2 \sin ^3 \theta}{2\pi c^2} d^2 \left|\frac{E_{T}}{E_0^+(n\frac{\omega}{c}\sin \theta)}\right|^2
\end{equation}
\normalsize
\noindent In the low frequency limit $\omega \ll \omega_p$, according to Eqs.~(\ref{eq.kz0})(\ref{eq.kz1})(\ref{eq.kz2}) and Eq.~(\ref{eq.E0p}), the $\theta$-dependence in $E_0^+$ only appears in $k_{z2}=\frac{\omega}{c} \sqrt{1-n^2 \sin^2 \theta}$ whereas $k_{z0}=\frac{\omega}{c} \sqrt{n^2-1}$ and $k_{z1}=i \omega_p/c$. The integrand in Eq.~(\ref{eq.RadImp}) is thus dominated by its value in the region close to the pole:
\small
\begin{eqnarray}\label{eq.pole}
k_{z2,c}&\simeq&\sqrt{-\frac{2\omega}{c}\delta k_{SPP}}\nonumber\\
&\simeq&\frac{1}{\beta \delta_p} \left(\frac{\omega}{\omega_p} \right)^2 \left\{\frac{\beta^2-1}{\beta}\frac{n^2}{\sqrt{n^2-1}} \left(\frac{\omega}{\omega_p} \right)+i \right\}
\end{eqnarray}
\normalsize
\noindent when $k_{z2}$ follows the contour $\frac{\omega}{c} \sqrt{1-n^2 \sin^2 \theta}$ with $\theta \in [0,\pi/2]$ (see inset of FIG.~\ref{figS9}):. Substituting Eqs.~(\ref{eq.E0p}) and (\ref{eq.EzT}) in Eq.~(\ref{eq.RadImp}), the lozentian approximation of the integrand leads to the radiation impedance in the low frequency limit $\omega \ll \omega_p$:
\small
\begin{equation}\label{eq.Restimat}
\mathcal{R}(\nu=\omega/2\pi)\simeq\frac{1}{\beta n^5} \left(\frac{d}{\delta_p} \right)^2 \left(\frac{\omega}{\omega_p} \right)^3  Z_{vac}
\end{equation}
\normalsize
\noindent Unlike the low frequency noise which is bonded by the $RC$ frequency cut-off ($1/(2\pi RC) \sim 100 \, \mathrm{kHz}$), the spectral power density at optical frequencies involves the radiation impedance $\mathcal{R}(\nu)$ which does not exhibit any high-frequency cut-off. This radiation impedance corresponds to a directed emission at angle $\theta_p$ with:
\small
\begin{equation}\label{eq.angle}
\sin \theta _p \simeq \frac{1}{n} \left\{1+\frac{1}{2\beta^2} \left(\frac{\omega}{\omega_p} \right)^2 \right\}
\end{equation}
\normalsize
\noindent  Note that the angle of emission is slightly greater than the angle of total internal reflection of the flat substrate which explains the role of the conical prism. The radiation impedance due to the coupling between the tunneling current and the electromagnetic field is estimated at $\mathcal{R}(\nu) \sim 0.5 \, \mathrm{m} \Omega$ for $\lambda = 1 \, \mu \mathrm{m}$. We can also estimate the radiation impedance due the coupling of the current in the electrodes but the screening factor $(n\omega/\omega_p)^2 \sim 10^{-2}$  of the electric field in the metal leads to $\mathcal{R}(\nu) \sim 1 \, \mu \Omega$ which disagrees with experiment.  Note that our calculation neglects the Drude dissipation compared to the plasmon leakage in the substrate.

\begin{figure}
\begin{center}
\includegraphics[width=0.8\linewidth]{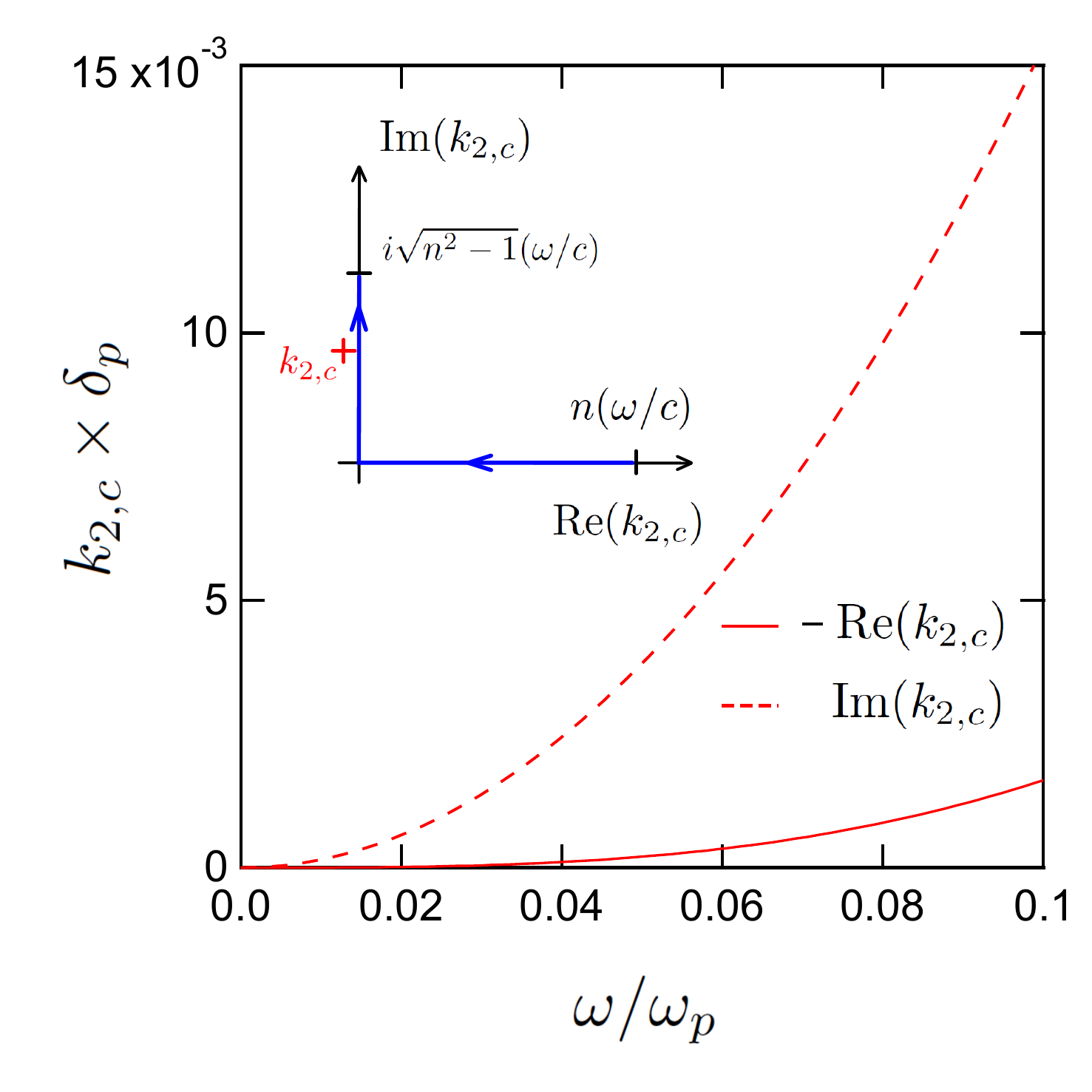}
\end{center}
\vspace{-0.5cm}
\caption{Frequency dependence of the pole $k_{z2,c}$ in the low frequency limit $\omega \ll \omega_p$. \textit{Inset}: illustration of the integration contour used to calculate the radiation impedance.\label{FigS9}}
\end{figure}

\subsection{Photon emission efficiency in a metallic tunnel junction}

\begin{figure}
\begin{center}
\includegraphics[width=0.8\linewidth]{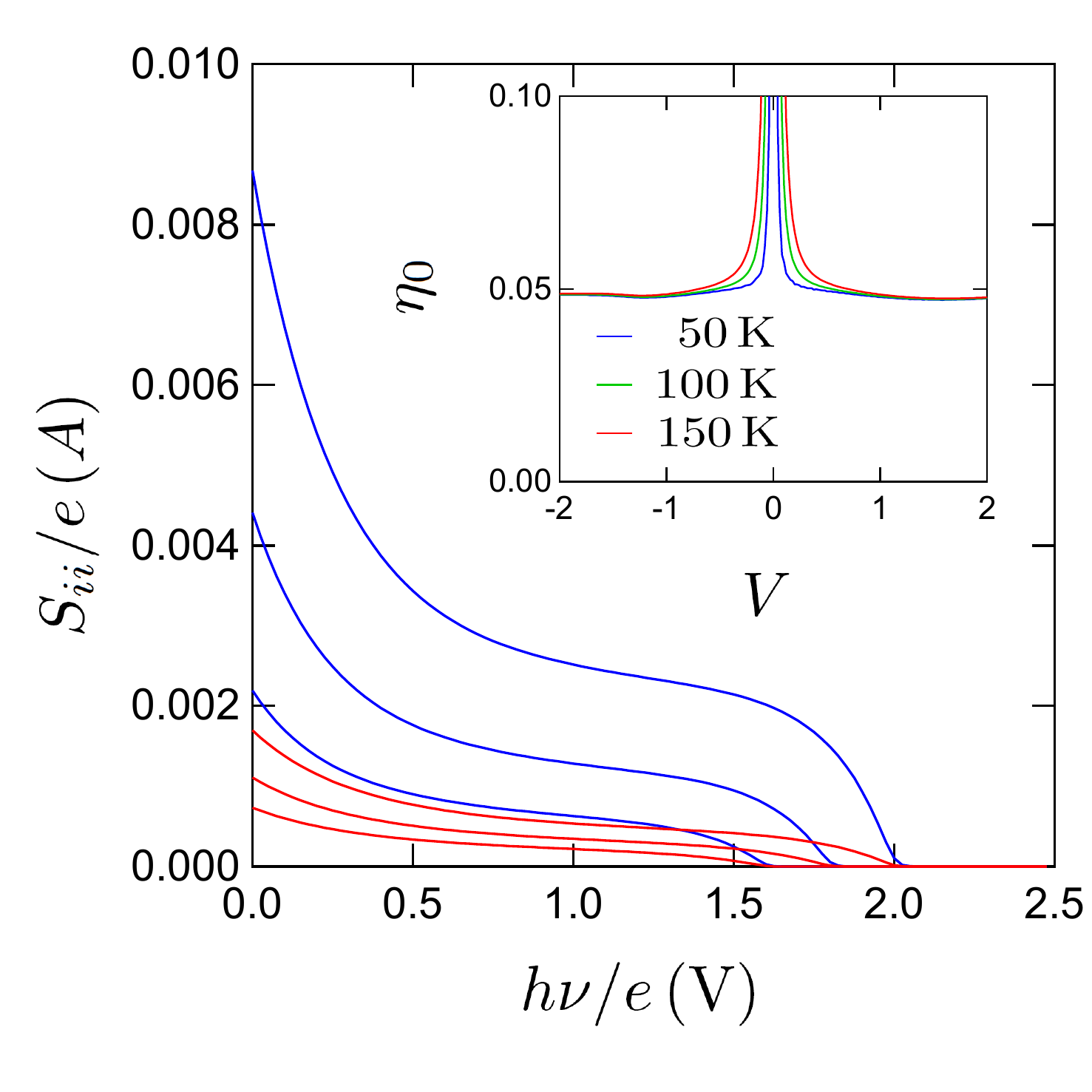}
\end{center}
\vspace{-0.5cm}
\caption{Theoretical current noise spectral density for different bias voltage. Red (\textit{resp.} Blue) lines corresponds to positive (\textit{resp.} negative) voltages $2$, $1.8$ and $1.6 \, \mathrm{V}$. \textit{Inset}: Voltage dependence of $\eta_0=\eta \times (R_K/\mathcal{R})$ for different temperatures in the low frequency limit $\omega \ll \omega_p$.\label{figS10}}
\end{figure}

The emission efficiency is usually defined by an electron to photon conversion rate:
\small
\begin{equation}
\frac{\int_{0}^{+\infty}P_{\nu}/(h\nu)d\nu}{I/e}
\end{equation}
\normalsize
\noindent However, as explained in the article, it is more relevant to define it with respect to the Joule power $P_J=V \times I$ dissipated in the tunnel junction:
\small
\begin{equation}\label{eq.eta}
\eta(V) = \frac{\int_{0}^{+\infty}P_{\nu}d\nu}{V\times I(V)}
\end{equation}
\normalsize
\noindent FIG.~\ref{figS10} shows the theoretical current noise spectral density for different bias voltage. The junction is characterized by the set of parameters $(U,\Delta U,d))$  defined in section \ref{section.param}. It exhibits the cross-over at $h\nu =eV$ as expected. It enables to numerically calculate the efficiency $\eta$ and demonstrate the relationship between the efficiency and the ratio $\mathcal{R}/R_K$ (see inset of FIG.~\ref{figS10}):
\small
\begin{equation}\label{eq.etaV}
\eta(V) \sim \eta_0 \frac{\mathcal{R}\left( \frac{eV}{h} \right)}{R_K}
\end{equation}
\normalsize
\noindent where $\eta_0 \simeq 0.047$ in the low frequency limit $\omega \ll \omega_p$ where the radiation impedance is given by Eq.~(\ref{eq.Restimat}). Note that $\eta_0$ is voltage dependent at low bias voltage giving rise to an increased efficiency. This is an artifact due to the black body radiation which are always emitting even at zero bias voltage. $\eta_0$ is a constant weakly dependent on the details of the barrier and depends mainly on the frequency dependence of the radiation impedance. Its numerical value is indeed close to $\eta_0=1/20$ find for a tunnel junction with constant transmission at zero temperature in the low frequency limit $\omega \ll \omega_p$:
\small
\begin{equation}\label{eq.eta0}
\eta =\frac{\int_{0}^{+\infty}\mathcal{R}(\nu) eI(V)\left(1-(h\nu/eV) \right) d\nu}{V\times I(V)}=\frac{1}{20}\frac{\mathcal{R}\left( \frac{eV}{h} \right)}{R_K}
\end{equation}
\normalsize
\noindent

\vspace{-5mm}
\bibliography{QcED}
\vspace{-6mm}

\end{document}